\documentclass[12pt]{article}
\pdfoutput=1
\usepackage{float}
\usepackage[final]{graphicx}
\usepackage{amssymb,amsmath}
\usepackage{commath}
\usepackage{epsfig}
\usepackage{epstopdf}
\usepackage{latexsym}
\usepackage{graphicx}
\usepackage{subfigure}
\usepackage{booktabs}
\usepackage{tkz-euclide}
\usepackage{makeidx}
\usepackage{cite}
\usepackage{bm}
\usepackage{geometry}
\usepackage{braket}
\usepackage[margin=20pt,small]{caption}

\usepackage[toc]{appendix}

\usepackage{tikz}
\usetikzlibrary{calc,decorations.markings,arrows.meta,shapes.misc,decorations.pathmorphing,calc,bending}

\usetikzlibrary{arrows.meta,shapes.misc,decorations.pathmorphing,calc,bending}

\tikzset{
  branch point/.style={cross out,draw=black,fill=none,minimum size=2*(#1-\pgflinewidth),inner sep=0pt,outer sep=0pt},
  branch point/.default=5
}
\tikzset{
  branch cut/.style={
    decorate,decoration=snake,
    to path={
      (\tikztostart) -- (\tikztotarget) \tikztonodes
    },
    }
  }

\usepackage{color}
\usepackage{datetime}

\ifpdf
\fi

\def\cO{{\cal O}}

\definecolor{cardinal}{rgb}{0.6,0,0}
\definecolor{darkgreen}{rgb}{0,0.5,0}
\definecolor{golden}{rgb}{0.92, 0.7, 0}
\definecolor{midnight}{rgb}{0, 0, 0.5}
\definecolor{darkblue}{rgb}{0.2, 0, 0.8}

\newcommand{\be}{\begin{equation}}
\newcommand{\ee}{\end{equation}}
\newcommand{\bea}{\begin{eqnarray}}
\newcommand{\eea}{\end{eqnarray}}

\begin{document}

\begin{titlepage}

\centerline{\Large \bf Out-of-Time-Order correlators in driven
conformal field theories}
\bigskip
\centerline{\bf Suchetan~Das$^{1}$, Bobby~Ezhuthachan$^2$,
Arnab~Kundu$^{3,4}$, } \centerline{\bf Somnath~Porey$^2$,
Baishali~Roy$^2$, K.~Sengupta$^5$}
\bigskip
\centerline{$^1$Department of Physics, Indian Institute of Technology Kanpur, Kanpur 208016, India.}
\bigskip
\centerline{$^2$Ramakrishna Mission Vivekananda Educational and Research Institute,}
\centerline{Belur Math, Howrah-711202, West Bengal, India.}
\bigskip
\centerline{$^3$ Saha Institute of Nuclear Physics, 1/AF, Bidhannagar, Kolkata 700064, India.}
\centerline{$^4$Homi Bhaba National Institute, Training School Complex, Anushaktinagar, Mumbai 400094, India.}
\bigskip
\centerline{$^5$ School of Physical Sciences, Indian Association for the Cultivation of Science,}
\centerline{2A and 2B Raja S.C.Mullick Road, Jadavpur, Kolkata-700032, West Bengal, India.}
\bigskip
\centerline{suchetan[at]iitk.ac.in,  bobby.phy[at]gm.rkmvu.ac.in, arnab.kundu[at]saha.ac.in}
\centerline{somnathhimu00[at]gm.rkmvu.ac.in, baishali.roy025[at]gm.rkmvu.ac.in, tpks[at]iacs.res.in }

\begin{abstract}

\noindent {We compute Out-of-Time-Order correlators
(OTOCs) for conformal field theories (CFTs) subjected to either
continuous or discrete periodic drive protocols. This is achieved
by an appropriate analytic continuation of the stroboscopic time.
After detailing the general structure, we perform explicit
calculations in large-$c$ CFTs where we find that OTOCs display an
exponential, an oscillatory and a power-law behaviour in the heating
phase, the non-heating phase and on the phase boundary,
respectively. In contrast to this, for the Ising CFT representing an
integrable model, OTOCs never display such exponential growth. This observation hints towards how OTOCs can demarcate between
integrable and chaotic CFT models subjected to a periodic drive. We further explore
properties of the light-cone which is characterized by the
corresponding butterfly velocity as well as the Lyapunov exponent.
Interestingly, as a consequence of the spatial inhomogeneity
introduced by the drive, the butterfly velocity, in these systems,
has an explicit dependence on the initial location of the operators.
We chart out the dependence of the Lyapunov exponent and the
butterfly velocities on the frequency and amplitude of the drive for
both protocols and discuss the fixed point structure which
differentiates such driven CFTs from their undriven counterparts.}

\end{abstract}

\newpage

%\tableofcontents

\end{titlepage}
\tableofcontents

%\newpage

\rule{\textwidth}{.5pt}\\

%%%%%%%%%%%%%%%%%%%%%%%%%
\numberwithin{equation}{section}
\section{Introduction}
%%%%%%%%%%%%%%%%%%%%%%%%%

Stroboscopic dynamics of periodically driven closed quantum systems
has been intensely studied in recent years
\cite{rev1,rev2,rev3,rev4,rev5, rev6,rev7,rev8}. The reason for such
intense activity in the field is two-fold. First, at least some
theoretical predictions emerging from such studies have received
support from experiments carried out on ultracold atoms in optical
lattices \cite{rev9,exp1,exp2,exp3,exp4}. Second, many of such
studies lead to understanding of several phenomena that have no
analogue in equilibrium quantum systems.

The properties of any driven quantum system is controlled by its
unitary evolution operator $U(t,0)$ given by $U(t,0)= {\mathcal
T}_t\{\exp[-i \int_0^t H(t') dt'/\hbar]\}$, where $H(t)$ is the
(time-dependent) Hamiltonian of the system, ${\mathcal T}_t$ denotes
time ordering, and $\hbar$ is the Planck's constant. For
periodically driven systems, characterized by a time period $T$,
this evolution operator, at times $t= nT$, where $n \in Z$,
satisfies \cite{rev6,rev8}
\begin{eqnarray}
U(nT,0) &=& {\mathcal T}_t \left\{\exp\left[-i n \int_0^{T} H(t')
dt'/\hbar\right]\right\} = \exp\left[-i n T H_F(T)/\hbar\right]
\label{fl1}
\end{eqnarray}
where $H_F(T)\equiv H_F$ is the Floquet Hamiltonian of the system.
Thus the Floquet Hamiltonian of the systems controls stroboscopic
dynamics of any periodically driven system. However, its exact
computation for a generic non-integrable system is challenging due
to time ordering involved in the definition of $U$; consequently, it
is customary to resort to perturbative methods for its computation.
For small $T$, the Magnus expansion constitutes one such
perturbative scheme \cite{magrev,rev8}; however, its prediction
starts to deviate from the exact result as one approaches the
intermediate frequency regime. In contrast, the Floquet perturbation
theory \cite{fpt1,fpt2,rev8}, provides a more accurate description
of such driven systems in the intermediate and small $T$ regime
provided that the drive amplitude is large \cite{rev8}. However, at
large $T$, there are no known reliable analytic scheme for computing
$H_F$; in this regime, one has to usually rely on exact numerics.

Out of the several possible protocols used to drive a system out of
equilibrium, periodic protocols which can be understood in terms of
their Floquet Hamiltonians have been studied most intensively in
recent years. The main reason for this focus is the presence of
several phenomena in such systems that are usually not found in
aperiodically driven systems. These include generating quantum
states with non-trivial topology \cite{topo1,topo2,topo3,topo4},
realization of Floquet time crystals \cite{tc1,tc2,tc3},
demonstration of dynamical localization
\cite{dloc1,dloc2,dloc3,dloc4} and dynamical freezing
\cite{df1,df2,df3,df4}, induction of dynamical phase transitions
\cite{dtran1,dtran2, dtran3, dtran4}, and tuning ergodicity of a
quantum many-body system using frequency of the drive
\cite{erg1,erg2}. More recently, systems with quasiperiodic and
aperiodic drives \cite{qd1,qd2,qd3,nd1} has also been studied in
this context. See also \cite{Nizami:2020agu} for a discussion on measures of quantum chaos in Floquet systems.

A class of such driven systems involves conformal field theories
(CFT) subjected to periodic drives. It is usually expected that
driving a CFT would lead to generation of a timescale which will in
turn spoil its conformal invariance and drive it away from its
conformal fixed point \cite{sdas1}. However, recently, it was shown
\cite{Wen:2018agb, Wen:2020wee, Fan:2020orx, Han:2020kwp,
Andersen:2020xvu, Das:2021gts} that this is not necessarily the
case; indeed it is possible to subject a CFT to a periodic drive
without spoiling the conformal symmetry of the problem. See
 \cite{Lapierre:2019rwj, Lapierre:2020roc, Lapierre:2020ftq} for more studies in Floquet CFT systems. As explained
in Sec.\ \ref{setup}, this can be most easily done using a CFT model
with sine square deformation whose Hamiltonian is valued in ${\rm
su}(1,1)$\cite{Okunishi:2016zat, Milsted:2017csn}. The action of the
drive in this case leads to an evolution operator
\begin{eqnarray}
U(T,0) &=& {\mathcal T}_t \left\{ \exp\left[= i \int_0^T dt
H(t)/\hbar\right] \right\} = \left( \begin{array}{cc} a & b
\\ c & d \end{array} \right)  \label{ueq}
\end{eqnarray}
where $ad-bc=1$ and the coefficients $a$, $b$, $c$, and $d$ depend
on the drive protocol. We note that $U$ is valued in ${\rm
SU}(1,1)$. Since this group is isomorphic to ${\rm SL}(2,R)$, this
allows us to write $U$ in imaginary time as
\begin{eqnarray}
U(T=i \tau,0) \equiv U_i &=& \left( \begin{array}{cc} a_i & b_i
\\ c_i & d_i \end{array} \right) \label{ueq2}
\end{eqnarray}
where $a_i d_i-b_i c_i=1$. The action of $U_i$ in the complex plane
generates a M\"{o}bius transformation $ z \to z_n= (a_i z+b_i)/(c_i
z+d_i)$. The evolution of the quantum state under $U_i$ gets translated into the dynamics of operators of the CFT, when we move to the Heisenberg picture. Thus the stroboscopic time-dependence of
any primary operator can be obtained using the relation
\begin{eqnarray}
U^{\dagger}(T,0) O(z, {\bar z}) U(T,0) &=& \left( \frac{\partial
z_n}{\partial z}\right)^h \left( \frac{\partial {\bar z_n}}{\partial
{\bar z}}\right)^{\bar h} O(z_n,{\bar z}_n) \label{opevoldef}
\end{eqnarray}
where one analytically continues to real time at the end of the
calculation and $(h, \bar h)$ denotes the conformal dimension of
$O$. Using this prescription, energy density, equal and unequal-time
correlation functions, and entanglement entropies of driven CFTs
have been computed in both cylindrical and strip geometries
\cite{Wen:2018agb, Wen:2020wee, Fan:2020orx, Han:2020kwp,
Andersen:2020xvu, Das:2021gts}. An interesting property found in all
these quantities is the emergence of spatial inhomogeneity due to
the drive which is usually not found in typical condensed matter
systems.
 However, out-of-time-order correlation function for such driven CFTs has not been
studied so far.

The Out-of-Time-Order correlators (OTOC) are known to provide a
diagnostic for scrambling which precedes thermalization in a typical
non-integrable quantum many-body system
\cite{Das:2019tga,Dolan:2000ut,Fitzpatrick:2014vua,Roberts:2014ifa,Das:2021qsd,Aramthottil:2021cov}.
Examples of such systems include large $N$ spin and bosonic models
\cite{otoc1,otoc2a,otoc2b}. For such models, which have
well-understood semi-classical limits, the early time behavior of
the OTOC constitutes an exponential growth and can be written as
\begin{eqnarray}
C(x,t) &=& {\rm Tr}\left[ e^{-\beta H/2} W(x,t) V(0,0) e^{-\beta
H/2} W(x,t) V(0,0) \right]  \sim e^{ \lambda_{\rm L}(t- x/v_{\rm B})}
\label{otocdef}
\end{eqnarray}
where $\beta=(k_B T_0)^{-1}$ is the inverse of the temperature $T_0$
and $k_B$ is the Boltzman constant. Note that at zero temperature,
the trace in Eq.\ \ref{otocdef} is replaced by expectation in the
ground state and this will be the limit which we shall be interested
in here. The quantity $\lambda_{\rm L}$ is the equivalent of the
Lyapunov exponent in a classical chaotic system; its inverse
$t_{\ast}=1/\lambda_{\rm L}$ provides the timescale for information
scrambling. It is well known that in thermal system $\lambda_{\rm
L}$ satisfies the bound $\lambda_{\rm L} \le 2\pi/(\beta \hbar) \sim
T_0$ \cite{Maldacena:2015waa}. In contrast, $v_{\rm B}$, the
butterfly velocity, is bounded only by the speed of light $c$ and
measures the speed with which local perturbations grow. It
constitutes an analogue of the Lieb-Robinson velocity
\cite{Lieb:1972wy} for information spreading in the present case.
Such OTOCs have been studied in several contexts both in condensed
matter physics ({\it e.g.}~\cite{Xu:2018dfp, Gharibyan:2018fax}),
quantum field theories ({\it e.g.}~\cite{Steinberg:2019uqb,
Gu:2021xaj}), and conformal field theories with AdS duals ({\it
e.g.}~\cite{Poojary:2018esz, Banerjee:2018twd, Malvimat:2021itk}).

In this work, we extend the studies of such OTOCs to periodically
driven CFTs.  One of the motivations of our present
study is to understand the efficacy of the OTOC as a diagnostic of
chaos, beyond thermal equilibrium\footnote{ This is
part of a larger goal to use OTOCs to study quantum chaos in systems
outside its original purview, ie:-  in QFTs without boundaries and
in thermal equilibrium. OTOCs in QFTs in the presence of a boundary
has been studied in \cite{Das:2019tga}. See also
\cite{Das:2021qsd},\cite{Aramthottil:2021cov}, where OTOCs are
studied in in another out of equilibrium setting-the CC state in a
quantum quench.}. Floquet CFTs provide us with a set-up where we
have analytic control to answer this question. We answer this
question in the affirmative and show that unlike lower point
function probes such as the entanglement entropy and expectation
value of the energy, the 4-pt OTOC can demarcate integrable and
chaotic CFT's in the heating phase.

One key difference of these systems from the undeformed CFT is that
they break both spatial translational as well as time translational
symmetry. Another key difference is the existence of fixed points of
the flow under the deformed Hamiltonian, unlike the undeformed one,
where there are no fixed points.  These differences leads to novel
consequences, which have no analogue in the undeformed thermal CFT
case. Our main findings are as follows:

\begin{itemize}
\item[1.] For the continuous drive case, we show that in a large $c$ CFT,
the OTOC as defined in (\ref{otocdef}), shows an exponential growth
 for sufficiently large $n$, smaller than the scrambling time in the heating phase, but only for a range
of values of $x$ bounded by the fixed points of the flow. These fixed points are characteristic of driven CFTs; they
have no analogue in CFTs in equilibrium. The Lyapunov exponent, in
this example, is a function of the drive parameters. In the
non-heating phase, there is no exponential growth for any value of
$x$.

\item[2.] In the same example, we show that the butterfly velocity is position dependent.
 This is a consequence of the lack of translational
invariance in these systems. Our definition of the
butterfly velocity is a natural generalization of the same in the
undeformed case. We show that the butterfly velocity in such driven
systems can be tuned using drive frequency.

\item[3.] We further generalize our study of the OTOCs to the case where the two operators ($V$ and $W$) are kept at arbitrary points ($x_1$) and ($x_2$), instead of ($x$) and ($0$). In the undeformed case, due to spatial homogeneity, one can always choose one of the operators to be at the origin, without any
loss of generality. This is not true in the driven CFT case, and as
a result the OTOC is naturally a function of the two positions of
the operators and not only a function of the relative distance
between the two. Interestingly, for sufficiently large $n$, the
expression for the OTOC factorizes as a result of which the explicit
expression of the butterfly velocity remains same as the case when
the operators are placed at $x$ and $0$.

\item[4.] Finally, we study OTOCs in the case of the discrete drive where again the two operators ($V$ and $W$) are kept at arbitrary point ($x_1$) and ($x_2$).  The qualitative features remain the same as before, even though the expressions for the butterfly velocity and the Lyapunov exponent are different.

\item[5.] For driven Ising CFT, we find that the OTOC does not show exponential behaviour even in the heating phase. We expect our result to be true for more general integrable CFTs. Thus the OTOC, unlike lower point functions, distinguishes between integrable and chaotic CFTs, under the periodic drive.
\end{itemize}

The plan of the rest of the work is as follows. After a quick review
of the dynamics of the driven CFT systems in the next section, we
present our analysis of the OTOCs in the section \ref{OTOC}. We
first discuss the OTOC in the case of the continuous drive in
section \ref{contotoc}. We present the explicit results in the case
of a large $c$ CFT first for the case where the operators are placed
at $x$ and $0$, and then later for the more general case where  the
positions are taken to be $x_1$ and $x_2$. In section
\ref{discotoc}, we study the case of a discretely driven CFT, with
two periods $T_1$ and $T_2$. Section $4$ is devoted to a brief discussion on the Holographic description of our results. In the appendix, we discuss the OTOC computation for the driven Ising CFT. We conclude with a discussion of our
results in section \ref{discuss}.

\numberwithin{equation}{section}
\section{Driven CFTs: The set-up}
\label{setup}

Driven many-body systems provide a useful set-up for studying
non-equilibrium dynamics.  Among the many driving protocols, a very
interesting class is realized by the so-called Floquet dynamics of
CFT, see {\it e.g.}~\cite{Wen:2018agb, Wen:2020wee, Han:2020kwp,
Andersen:2020xvu, Das:2021gts}. This class of systems, despite a
non-trivial driving, preserves conformal symmetry and allows for an
analytical control over the dynamics, see {\it
e.g.}~\cite{Okunishi:2016zat, Milsted:2017csn}. In this protocol,
the drive is executed by independently controlling, the left (${\bf
T}(x-t)$)and right moving components $\bar{{\bf T}}(x+t)$ of the
energy-momentum density $({\bf T}_{00}(x,t))$ of the CFT. The
Hamiltonian, $H$, of the driven CFT, at $t=0$ is given by
\begin{equation}
H = \int dx \; (f(x) {\bf T}(x) + g(x) \bar{{\bf T}}(x)) \ .
\end{equation}

Here ${\bf T}_{00}= \frac{1}{2\pi}({ \bf T}+ \bar{{\bf T}})$, while
$f(x)$ and $g(x)$ represent the externally controlled functions,
which are both unity for an undriven CFT. The drive is accomplished
by changing these functions periodically, quasi-periodically or
randomly, either in discrete steps or continuously.  In this note,
we will be concerned only with periodically driven CFTs on a ring
under  both the continuous as well as the discrete drive. Under the
periodic drive protocol, evolution of the system is periodic with a
period $T$, so that the total evolution happens in steps($n$) of
$T$, {\it i.e.} $t=nT$. In the discrete drive case, the evolution within
this time period $T$ is updated discretely ({\it i.e.} $ T = T_1 + T_2 +
T_3 +... +T_k$). During each of these time intervals $T_j$, the
Hamiltonian is changed externally, for example, by updating the
functions $f(x)$ and $g(x)$. Consequently, the Hamiltonian for the
time $T_j$ may be parameterized by an integer ($H(j)$). In contrast,
for the continuous drive case, it is updated continuously throughout
the time period $T$.  We will now review, in more detail, the basic
setup for these two cases.

On a ring, the functions $f(x)$ and $g(x)$ have to obey periodic
boundary conditions, and thus admit a Fourier series expansion in
terms of its modes. This fact then allows one to express the
Hamiltonian as a linear sum of the Virasoro generators. Thus for
generic functions $f$ and $g$ and a discrete drive (say), the
Hamiltonian at the $j^{th}$ iteration, takes the form

\begin{equation}
H^{(j)} = \sum_{n} \Big(a^{(j)}_n L_{n} + b^{(j)}_{n}\bar{L}_{n}\Big) \ .
\end{equation}

Here $L_n$ and $\bar{L}_n$ generate two independent Virasoro algebras
\begin{eqnarray}
&&\left[ L_m, L_n \right] = \left( m-n\right) L_{m+n} + \frac{c}{12} m\left( m^2 -1\right) \delta_{m+n, 0} \ , \nonumber \\
&&\left[ \bar{L}_m, \bar{L}_n \right] = \left( m-n\right) \bar{L}_{m+n} + \frac{c}{12} m\left( m^2 -1\right) \delta_{m+n, 0}
\end{eqnarray}
and are related to the chiral and anti-chiral stress tensor
components in the usual way.
\begin{eqnarray}
 L_{n} = \frac{L}{2\pi} \int_0^{2\pi} dx e^{i\pi n x} {\bf T}_{00}(x) \ , \; \; \bar{L}_{n} = \frac{L}{2\pi} \int_0^{2\pi} dx e^{-i\pi n x} {\bf T}_{00}(x)
\end{eqnarray}
with $n\in {\mathbb Z}$, $x\in [0,2\pi]$ is an angular coordinate on
the $S^1$ and $L$ is the circumference of the ring.

For general $f$ and $g$, the sum is over all the modes and hence the
Hamiltonian dynamics is complicated, due to the presence of these
infinite number of generators. One can simplify the dynamics, by
restricting the form of the functions $f$ and $g$, so that the
Hamiltonian has contributions only from generators of a sub-algebra
of the Virasoro algebra. In this work, we will restrict ourselves to
the case when $f=g$ and the Hamiltonian is built out of the
generators of the diagonal ${\mathfrak sl}(2, {\mathbb R})$
sub-algebra of the global $sl(2,\mathbb{C})$, generated by
$\{L_0+\bar{L}_0, L_{\pm 1}+\bar{L}_{\pm 1}\}$\footnote{Note that,
by constructing a Hamiltonian from linear combinations of $\{L_0,
L_{\pm m}\}$ also leads to a similar control on the dynamics. The
corresponding algebra is ${\mathfrak sl}^{(m)}(2, {\mathbb R})$. The
corresponding group is isomorphic to a $m$-fold cover of the SL$(2,
{\mathbb R})$ group. See {\it e.g.}~\cite{Wen:2020wee, Fan:2020orx}
for an analysis of this more general case.}.

In this case, the evolution operator at each step is the usual
$sl(2,\mathbb{R})$ transformation, and consequently the full time
evolution is given by combining the conformal transformation at each
step, which is again an $sl(2,\mathbb{R})$ transformation. To work
out the full dynamics, the strategy used is as follows. We first
Wick rotate the Lorentzian time to Euclidean time and then map the
cylinder to the plane, using: $w = \frac{L}{2\pi} \log z$. On the
plane, once we obtain the expressions of the combined conformal
transformation, we Wick rotate back to Lorentzian time, to obtain
the corresponding transformation in real time.

\subsection{ Time evolution in Continuously Driven CFTs}\label{sec cont}

The Hamiltonian of the driven system in this case, is given by
\begin{eqnarray}
H(t') = \frac{2\pi}{L} \left[ f(t') L_0 + \frac{1}{2}f_1(t')  \left( L_1 + L_{-1}\right) \right] + {\rm anti-holomorphic}\ ,  \label{floham}
\end{eqnarray}
where $\{f(t'), f_1(t')\}$ are real-valued functions that encodes the drive protocol and $t'$ is an auxiliary parameter
that describes the function-space. Setting $f_1(t') =0$ and $f(t') =
1$ yields us the standard (undriven) CFT Hamiltonian on the
cylinder. Given (\ref{floham}), the corresponding evolution operator
can be calculated. This yields:
\begin{eqnarray}
&& U\left(T, 0 \right) = {\rm exp} \left[ - i \int_0^T H(t) dt \right]  \quad \implies \quad U(T, 0) =  \left[ {\begin{array}{cc}
   a & b \\
   c & d \\
  \end{array} } \right] \ . \label{evosl2r}
\end{eqnarray}
Since $H(t)$ only contains $L_{0,\pm 1}$, the evolution operator
defined above is an element of the SL$(2, {\mathbb R})\simeq {\rm
SU}(1,1)$ group: $ad - bc =1$. Here, $T$ is a parameter associated
to the drive protocol, which is real-valued. The evolution is
step-wise: we imagine driving the system with a Hamiltonian $H(t)$
for an interval of $T$, for $n\in {\mathbb Z}$ number of cycles.
This $n$ plays the role of stroboscopic time. For a given $T$, and
an integer $n$, $nT$ measures the time. Here, the Hamiltonian $H(t)$ is determined by choosing a drive
protocol data $\{f(t), f_1(t)\}$. Correspondingly, in the Heisenberg
picture, operator evolutions for stroboscopic times $t= nT$, are
simply given by
\begin{eqnarray}
&& \cO(x,t) = U^\dagger(t,0) \cO(x, 0) U(t,0),  \quad U(t) =
U^n(T,0), \quad U(T,0) = e^{- i \int_0^{T}H(t') dt' }.
\end{eqnarray}

Above, we have introduced a notion of time, denoted by $t$.
Clearly, $n \in {\mathbb Z}$, but $t \in {\mathbb R}$.

The map from the cylinder coordinates($w$) to the plane
coordinates($z$), is given by $w = \frac{L}{2\pi} \log z$. Thus on
the plane, we have (Eq.\ \ref{ueq2}):
\begin{eqnarray}
U_i^\dagger \cO(z, \bar{z}) U_i = \left( \frac{\partial z' }{\partial z} \right)^h \left( \frac{\partial \bar{z'} }{\partial z} \right)^{\bar{h}} \cO(z', \bar{z'}) \ ,
\end{eqnarray}
As explained earlier, $U_i \in {\rm SL}(2, {\mathbb R})$. On the plane, this acts as:
\begin{eqnarray}
z' = \frac{a_i z + b_i}{c_i z + d_i} \ , \quad {\rm with} \quad \left[ {\begin{array}{cc}
   a_i & b_i \\
   c_i & d_i \\
  \end{array} } \right] \in {\rm SL}(2, {\mathbb R}) \ .
\end{eqnarray}
Correspondingly, the complete evolution of the Heisenberg operator is determined by a combined conformal transformation:
\begin{eqnarray}
&& \left( \prod_i U_i^\dagger \right)  \cO(z, \bar{z}) \left( \prod_i U_i \right) = \left( \frac{\partial z_n }{\partial z} \right)^h \left( \frac{\partial \bar{z}_n }{\partial z} \right)^{\bar{h}} \cO(z_n, \bar{z}_n) \\
&& {\rm with} \quad z_n = \frac{a_n z_{n-1} + b_n}{c_n z_{n-1} + d_n} \ , \ldots \quad z_1 = \frac{a_1 z + b_1}{c_1 z + d_1} \ .\label{sl2rn}
\end{eqnarray}
The last line above defines a recursion relation that determines
$z_n$ at the $n^{\rm th}$-step, starting from the initial assignment
of $z$. A similar transformation holds for the anti-holomorphic part
as well.

For explicit calculations, we collect some important results of
\cite{Das:2021gts}. The evolution operator in (\ref{evosl2r}) can be
written as $U(T,0)=e^{- i H_F T}$, where $H_F = p(T) \sigma_z + i
q(T) \sigma_y$. Here $\sigma$'s are the Pauli matrices
and $p(T)$ and $q(T)$ can either be numerically exactly determined
or computed analytically within FPT \cite{Das:2021gts}. The former
is obtained by dividing $T$ into $N = T/\delta T$ steps, within each
of which the Hamiltonian in (\ref{floham}) remains approximately
constant. The total $U(T,0)$ is constructed by multiplying the
resulting step-wise evolution operators.

Now, following \cite{Das:2021gts}, the drive protocol data is given
by $f_1(t)=1$ and
\begin{eqnarray}
f(t) = f_0 \cos\left( \omega_D t \right) + \delta f \ .
\end{eqnarray}
Here the drive amplitude $f_0 \gg \delta f$, where $\delta f$ is
 the static (DC) component of the drive. In the
regime $\omega_D \ge \delta f, 1$, the following parameter
characterizes the dynamics:
\begin{eqnarray}
\alpha = \sum_{n=-\infty}^{\infty} J_n\left( \frac{2f_0 \pi}{L \omega_D}\right)  \frac{T}{n\pi + \pi\frac{\delta f T}{L}} \ ,
\end{eqnarray}
where $J_n$ are the Bessel functions of the first kind. In
particular, $\alpha^2>1$ ($\alpha^2<1$) corresponds to the heating
(non-heating) phase while $\alpha^2 =1$ describes the phase
boundary. We note that in the limit $\omega_D L \gg 2 f_0 \pi$, one
simply obtains: $\alpha \simeq L/(\pi \delta f)$,\footnote{One makes
use of the asymptotic behaviour of Bessel function of the first
kind:
\begin{eqnarray}
J_n(x) \approx \frac{1}{\Gamma(n+1)} \left( \frac{x}{2}\right)^n \ , \quad {\rm for} \quad 0< x \ll \sqrt{n+1} \ ,
\end{eqnarray}
along with $J_{-n}(x) = (-1)^n J_n(x)$.} which is independent of the
drive amplitude $f_0$ and the drive frequency $\omega_D$. Setting
further $L\gg \delta f$, we get $\alpha \gg 1$. For a large drive
frequency, we obtain a heating phase for small $\delta f$; the
transition to a non-heating phase occurs when $\delta f = \pi/L$. In
the opposite limit of small frequencies, no such simplification
occurs and $\alpha$ depends on all parameters in the drive protocol.

The last ingredient that we need is how the drive protocol data are
related to the SL$(2, {\mathbb R})$ transformations in
(\ref{sl2rn}).  For $\alpha^2>1$, the analytical
expressions of these quantities follow from those of $p(T)$ and
$q(T)$ computed using FPT and are given by\cite{Das:2021gts}
\begin{eqnarray}
&& a_n = \cosh\left( n \theta \right) - i \frac{1}{\sqrt{\alpha^2-1}} \sinh \left( n \theta\right) \ , \quad d_n = a_n^* \ ,\quad \theta= s \sqrt{\alpha^2 -1} \\
&& b_n = \frac{i \alpha}{\sqrt{\alpha^2-1}} \sinh\left( n
\theta\right) \ , \quad c_n = b_n^* \ , \quad s = \cos^{-1} \left[
\cos\left( \frac{\delta f T \pi}{L}\right) \right] \ .
\label{abcddef}
\end{eqnarray}
We note that the unitarity of the evolution matrix in the heating
phase is preserved for $\delta f T/L = -n$ (for $n<0$) for which
$\alpha$ diverges.

For $\alpha^2<1$, similar relations hold, except that $\cosh \to
\cos$ and $\sinh \to \sin$. The $\alpha^2=1$ boundary needs a
separate analyses. In this case, we obtain\cite{Das:2021gts}:
\begin{eqnarray}
&& a_n = 1 - i n s \ , \quad d_n = a_n^* \ , \\
&& b_n = - i  n s \ , \quad c_n = b_n^* \ .
\end{eqnarray}
We shall use these results for computation of OTOC in Sec.\
\ref{OTOC}.

%%%%%%%%%%%%%%%%%%%%%%%
\subsection{Time evolution in Discretely Driven CFTs}
%%%%%%%%%%%%%%%%%%%%%%%

We now consider a discrete drive, similar to the ones considered in
\cite{Wen:2020wee}.  We shall restrict ourselves here to a two-step
protocol for which the Hamiltonian is given by :
\begin{eqnarray}
H(t) &=& H_1 \quad {\rm for} \quad T_1 \ , \\
&=& H_2 \quad {\rm for} \quad T_2 \ ,
\end{eqnarray}\label{discrete}
where $T_{1,2}$ are the corresponding time intervals with
$T_1+T_2=T$. Using the notations of \cite{Wen:2020wee}, the
explicit Hamiltonians are given by
\begin{eqnarray}
&& H_\phi = \int_0^L {\bf T}_{00} dx \left( 1 - \tanh (2\phi) \cos \left( \frac{2\pi x}{L}\right) \right) \ ,\\
&& H_1 = H_{\phi=0} \ , \quad H_2 = H_{\phi \not =0} \ .
\end{eqnarray}
The full Hamiltonian can be written in terms of $\{L_0, L_{\pm 1}\}$
that generate the SL$(2,{\mathbb R})$ symmetry\cite{Wen:2020wee}.
Thus, the corresponding time evolution is governed by a set of
conformal transformations, corresponding to $H_1$ and $H_2$ with
time-intervals $T_1$ and $T_2$, {\it ad infinitum}.

The explicit M\"{o}bius-transformation, for a single step, is given by
\begin{eqnarray}
&& M_\phi \left( H_\phi, T_\phi \right)  = \left[ {\begin{array}{cc}
   a & b \\
   b^* & a^* \\
  \end{array} } \right] \ , \\
 && {\rm with } \quad a = \cos \left( \frac{\pi T_\phi}{L_{\rm eff}} \right) + i \cosh(2\phi) \sin \left( \frac{\pi T_\phi}{L_{\rm eff}} \right) \ , \\
 && b = - i \sinh (2\phi) \sin \left( \frac{\pi T_\phi}{L_{\rm eff}} \right)    \ , \quad L_{\rm eff} = L \cosh(2\phi) \ .
\end{eqnarray}
We intend to explore the heating-phase, which occurs for:
$\left(T_1, T_2 \right) = \left(L/2, L_{\rm eff}/2 \right)$. The
$n^{\rm th}$-evolution is obtained by multiplying products of
$\left(M_0 M_1\right) $, $n$ times. This yields
\begin{eqnarray}
U(nT,0) &=& \left( M_0 M_1\right)^n = \left(-1\right)^n \left(
{\begin{array}{cc}
   \cosh(2n\phi) & -\sinh(2n\phi) \\
   -\sinh(2n\phi) & \cosh(2n\phi) \\
  \end{array} } \right) \ .
\end{eqnarray}
We shall use this result in Sec.\ \ref{OTOC}.

%%%%%%%%%%%%%%%%%%%%%%%
\numberwithin{equation}{section}
\section{Correlation functions in Driven CFTs}\label{OTOC}
%%%%%%%%%%%%%%%%%%%%%%%

Correlation functions are the primary observables in any QFT. In the
context of driven CFTs, one is interested in the growth of the
energy density and entanglement entropy of a region with time. In
particular, different phases found in Floquet CFTs are classified by
the qualitatively different temporal dependence of these quantities.
For instance, in the ``heating phase", the energy grows exponentially
with $n$- the stroboscopic time, while the entanglement entropy
grows linearly. While the energy density is given by the one-point
function of the stress tensor in the appropriate state, the
computation of the entanglement entropy involves an equal time
two-point function of a specific ``twist operator", which is a
primary operator in the CFT.  In this note, we study four point
functions in these theories. In particular, we are interested in the
computation of the 4-point OTOC function in the vacuum state. These correlation
functions are known to be good probes of quantum chaos at early
times (times much smaller than the scrambling time).

We begin with a review of  OTOC in the undeformed 2D
CFT. The normalized $4$-point Euclidean correlator, of the
following form:
\begin{eqnarray}
&& \frac{\left \langle W(z_1, \bar{z}_1) W(z_2, \bar{z}_2) V(z_3, \bar{z}_3) V(z_4, \bar{z}_4)\right \rangle}{\langle W(z_1, \bar{z}_1) W(z_2, \bar{z}_2)\rangle \langle  V(z_3, \bar{z}_3) V(z_4, \bar{z}_4)\rangle } =  F\left( \eta, \bar{\eta} \right) \ , \\
&& \eta = \frac{z_{12} z_{34} }{z_{13} z_{24} } \ , \quad \bar{\eta} = \frac{\bar{z}_{12} \bar{z}_{34} }{\bar{z}_{13} \bar{z}_{24} } \ .
\end{eqnarray}
Here $F\left( \eta, \bar{\eta} \right)$ is an undetermined function
which depends on the dynamics of the corresponding CFT. The
dynamical function $F$ can be expanded in the basis of global
conformal blocks\cite{Dolan:2000ut}, which are given by
Hypergeometric functions, with coefficients that are determined by
the OPE-coefficients of the corresponding CFT. It is the latter that
encodes details of the CFT-dynamics. The Lorentzian correlators are
obtained by an analytic continuation in the complex time plane using
the $i\epsilon$ prescription.  To be specific, if we
define $z_i = x_i +i\tau_i$, where $\tau_i$ is the complex time ({\it i.e.} after continuing the time to the complex domain), then the
$i\epsilon$ prescription dictates that we take $\tau_i = it_i
+\epsilon_i$ in the limit $\epsilon_i\rightarrow 0$, where $t_i$ is
the Lorentzian time. The different possible ways in which the limits
can be taken correspond, in the Lorentzian theory, to the different
possible ordering of operators in the correlation functions. As an
example, taking the following ordering of limits ($\epsilon_1 >
\epsilon_3 > \epsilon_2
> \epsilon_4\rightarrow 0$), we get the following Lorentzian
correlation function $[\left \langle W(z_1, \bar{z}_1) V(z_3,
\bar{z}_3) W(z_2, \bar{z}_2) V(z_4, \bar{z}_4)\right
\rangle]/[\langle W(z_1, \bar{z}_1) W(z_2, \bar{z}_2)\rangle \langle
V(z_3, \bar{z}_3) V(z_4, \bar{z}_4)\rangle]$.  Since the $z_i$s
appear in the correlation function through a single complex number,
the cross ratio $\eta$, the analytic continuation happens
effectively in the complex $\eta$ plane. It is well-known that for a
2D CFT on the plane, the function $F(\eta)$ has branch cuts along
the real line from $\eta=1$ to $\eta =\infty$. The behavior of the
real-time correlation function depends crucially on  whether $\eta$
crosses this branch cut during the analytic continuation to real
time.

In the study of quantum chaos, the commutator square of two
observables  ($-\langle [V(x_1, t), W(x, (t')]^2\rangle_{\beta}$) in
the thermal state has been proposed to be a good measure of early
time chaos. Physically this measures the effect of a perturbation
$W$ at time $t'$ on the measurement of an observable $V$ at time
$t$. The thermal correlation
functions naturally live on a cylinder, with $\beta$ being the
radius of the circle. Using space time translation invariance, one
can set, without loss of generality, $(x_1=0)$, $(t >0)$ and
$(t'=0)$. In 2D CFTs, one can map
this geometry to the plane via an appropriate conformal map. Thus
effectively, with this map, we are computing vacuum correlation
functions on the plane of the type described above. When studying
quantum chaos, we are interested in the behaviour of the commutator
square, or equivalently, the 4-pt correlation functions at late
times $(t>> x)$. It can be shown that as we take $t$ from small
values to late times, the cross ratio $\eta$ transverses a closed
path trajectory in the complex time plane. It starts from very close
to the origin, crosses the real line at $t=x$, and finally again
takes a very small value at late times. In particular, with this
configuration, in 2D CFT in the thermal state, $\eta$ crosses the
real line at precisely $\eta
=\epsilon_{12}\epsilon_{34}/(\epsilon_{13}\epsilon_{24})$ (where
$\epsilon_{ij}= \epsilon_i-\epsilon_j$) in the
$\epsilon_i\rightarrow 0$ limit. It is easy to see that this is
greater than one for the out-of-time-ordered (OTO) configurations
and less than one for time-ordered (TO) configurations. Thus for OTO
configurations, $\eta$ crosses the branch cut during the analytic
continuation, while it does not do so for TO configurations. In a
quantum chaotic CFT, this fact results in a exponential growth of
the OTOC at late times. However, to see this explicitly one has to
work with theories where the explicit form of the function $F(\eta)$
is known. A prime example of this is the OTO in a ``large-$c$" CFT,
of two scalar Hermitian operators- ``the heavy operator" $W$ and ``the
light operator" $V$.\footnote{Here heavy and light refer to the fact
that the conformal dimension of $W$ scales linearly with $c$ in the
large-$c$ limit , {\it i.e.} $\frac{h_w}{c}$ is finite,  while
$\frac{h_v}{c}\rightarrow 0$ in the same limit. In these theories
and for this choice of operators, one can explicitly carry out the
analytic continuation in $\eta$ to get the exponential temporal
growth in the OTO in the thermal state at late times.}

We will now set up the analogous computation for these normalized
4-pt correlation functions, in the periodically driven CFTs, in the
vacuum state. Given the fact that the driven Hamiltonian we consider
here really generates an $sl(2,\mathbb{R})$ transformation under
which the vacuum state is actually invariant, one might not expect
any non trivial time evolution of these correlation functions in the
vacuum. For instance the energy density and entanglement entropy,
will be time independent in the vacuum state. However, this is true
only for equal time correlation functions.  Unequal time correlation
functions, either time-ordered or out-of-time ordered such as the
one considered in this work, display non-trivial dynamics. This
feature follows from the fact that they carry information about the
time-dependence of the Hamiltonian through the dynamics of the
operators in the Heisenberg picture.

\subsection{4pt OTOC in the continuously driven CFTs}\label{contotoc}
%%%%%%%%%%%%%%%%%%%%%%%%%%%%
\subsubsection{A Simple $4$-point OTOC}
%%%%%%%%%%%%%%%%%%%%%%%%%%%%

We begin by considering a correlator of the form $\left\langle W W V
V\right \rangle$, in which both $W$'s are placed at $x=0$ and the
$V$'s are placed at $x$. The initial time, for all the operators is
chosen to be zero. The corresponding arrangement, on the plane, is
given by:
\begin{eqnarray} \label{z1234in}
z_1 = 1 \ , \quad z_2 =1 \ , \quad z_3 = e^{\frac{2\pi ix}{L}}= z_ 4 \ ,
\end{eqnarray}
which also determines the corresponding complex-conjugates. As
described in the previous section, time evolution will be realized
as the stroboscopic time $n$ in conformal transformations on the
co-ordinates and therefore the operators located on them. For our
purpose, the two $W$ operators placed initially at the points
$\{z_1, z_2\}$ are evolved to time $n$ while the two $V$ operators
placed initially at the points $\{z_3, z_4\}$ do not undergo any
time evolution.  To impose operator ordering, we will shift in
euclidean time $n\theta +i\epsilon_i$, where the subscript refers to
the  $i^{\rm th}$ position
$z_i$. This has to be done for all operator positions. For this
purpose, it is useful to first consider time evolution of the $V$
operator by $m$ steps, do the $i\epsilon$ shift in $m\theta$ and
then put $m=0$. In addition to ensuring operator ordering, the
$i\epsilon$ prescription also regulates the ``contact divergence'' of
the correlation functions which arises from the fact that  the two $V's$ (as also
the two $W's$) were initially located at the same point.

After $n$-iteration of conformal transformations on $\{z_1,z_2\}$ and $m$-iteration of the same on $\{z_3,z_4\}$, one obtains:
\begin{eqnarray}
&& z_{1n} = \frac{a_n z_1+ b_n}{c_n z_1 + d_n} \ , \quad z_{2n} = \frac{a_n z_2+ b_n}{c_n z_2 + d_n} \ ,  \\
&& z_{3m} = \frac{a_m z_3+ b_m}{c_m z_3 + d_m} \ , \quad z_{4m} = \frac{a_m z_4+ b_m}{c_m z_4 + d_m} \label{etam} ,\\
&& \eta = \frac{\left( z_{1n}- z_{2n}\right) \left( z_{3m}- z_{4m}\right)}{ \left( z_{1n}- z_{3m}\right) \left( z_{2n}- z_{4m}\right) } \ . \label{eta}
\end{eqnarray}
But for the $\epsilon$ regulator, the numerator in the expression
for $\eta$ would be zero always and hence we expect $\eta$ to be
proportional to $\mathcal{O}(\epsilon^2)$. So $\eta$ would be finite
iff the denominator also is of the same order in $\epsilon$. This
happens when $z_{1n} - z_{3} \sim \mathcal{O}(\epsilon)$ and $z_{2n} -z_{4}\sim
\mathcal{O}(\epsilon)$.

The situation is similar to the underformed CFT case, where again
the expression for $\eta$ is finite when $(z_1{(t,0)} \sim
z_3(0,x))$. In that case, this happens when $(t\sim x)$. It is also
precisely at ($t=x$), that the analytically continued curve
$\eta(t)$ crosses the real axis in the $\eta$ complex plane. This
will turn out to be true for our case as well.

 We first analyze the case of continuous protocols
for which the coefficients $(a_n, b_n, c_n, d_n)$ are given by Eq.\
\ref{abcddef} for $\alpha^2>1$. In this case, one can work out the condition for $z_{1n}=z_3$. In the heating
phase, this turns out to be
\begin{eqnarray}
\tan \left( \frac{\pi x}{L}\right) &=&
\sqrt{\frac{\alpha-1}{\alpha+1}} \tanh(n\theta) \ .  \label{cond1}
\end{eqnarray}

Substituting this back into the expression of $\eta$, then yields:
\begin{eqnarray}
 \eta = \frac{\epsilon_{12} \epsilon_{34}}{\epsilon_{13}\epsilon_{24}} \implies \quad \eta < 1 \, \, {\rm with} \, \, \epsilon_1>\epsilon_2>\epsilon_3>\epsilon_4 \ , \quad \eta> 1 \, \, {\rm with} \, \,  \epsilon_1>\epsilon_3>\epsilon_2>\epsilon_4 \ . \label{realcut}
\end{eqnarray}
The above holds for values of $\{n, x\}$ satisfying (\ref{cond1}).
 We note here that initially ($n= 0$), $\eta$ vanishes
as $\epsilon\to 0$. Similarly, in the limit of large $n$, the
denominator is dominated by $e^{n\theta}$ and therefore $\eta \sim
e^{-2n\theta} \sim \cO(\epsilon^2) \to 0$.  Thus we can conclude
that for this arrangement of $\epsilon_i$'s, which corresponds to
the OTO configuration, $\eta$ starts of from near the origin at
$n=0$ and then crosses the real axis for a value greater than one,
and therefore crosses the branch cut in the complex plane, before
going to zero from the second sheet. It's interesting to note that
despite the difference in the dynamics, the nature of the analytic
continuation is very similar to the case of the undeformed CFT in
thermal equilibrium, where too the $\eta(t)$ curve crosses the real
axis at precisely the value given in (\ref{realcut}) and goes to
zero for large and small times.

There are however some important differences as well, which arise
due to the fact that the drive explicitly breaks time-translation
invariance as well as spatial homogeneity. This breakdown leaves an
imprint on lower point functions as can be explicitly seen in the
expectation values of one-point functions. For example, the relation
\begin{eqnarray}
\frac{2\pi x}{L} = \pi \pm \cos^{-1} \left( \frac{1}{\alpha}\right)  \ , \label{energypeak}
\end{eqnarray}
determines the positions of the energy peaks ({\it i.e.}~positions
where the one-point function of the stress tensor maximizes) that is
obtained from the one-point function of the
stress-tensor\cite{Das:2021gts}.

These spatial and time translation non-invariance also leads to key
differences at the level of the four point functions.  First, the
stroboscopic time $n$ is discrete, unlike the real continuous time
in the thermal CFT example. Thus the analytical continuation
produces a discrete set of points instead of a continuous curve.
Therefore, there may not be any integer $n$ solution to Eq.\
\ref{cond1}. This means that under the M\"{o}bius evolution, for a
particular value of $n$, $\eta(n)$ is a point above the real axis
({\it i.e.} in the upper half of the complex $\eta$ plane) while
for the next value ($n+1$), it's below the real axis (in the lower
half plane), without touching the real axis for any integer $n$. In
such a case, it becomes difficult to conclude whether after crossing
over, $\eta$ lies in the first or second Riemann sheet. We will use
Eq.\ \ref{cond1} as a criteria to decide that. Second, unlike the
equation $t=x$, which always has a solution for any $x$, Eq.\
\ref{cond1} has solutions only for a range of values of the initial
position $x$. This range is: $0 \le \pi x/L \le \tan^{-1}\left(
\sqrt{\frac{\alpha-1}{\alpha+1}} \right) $.

To complete the discussion on kinematics, let us confirm that if
$\eta$ crosses the branch-cut, then $\bar{\eta}$ does not, and
vice-versa. The crucial point is that both of them
should not simultaneously cross the branch-cut, to yield a
non-trivial dynamics in the analytically continued correlator. The
corresponding condition for the $\bar{\eta}(n)$ to cross the real
axis is given by
\begin{eqnarray}
 \tan \left( \frac{\pi x}{L}\right)  = - \sqrt{\frac{\alpha-1}{\alpha+1}} \tanh(n\theta) \ .
 \label{cond2}
\end{eqnarray}

Similar to (\ref{cond1}),  Eq.\ \ref{cond2} admit solutions in the
range, [$0 \ge \pi x /L \ge - \tan^{-1}(
\sqrt{\frac{\alpha-1}{\alpha+1}})$]. Taken together, the condition
that either $\eta$ or $\bar{\eta}$ crosses the branch cut can only
be satisfied for
\begin{eqnarray}
\left[- \tan^{-1} \left( \sqrt{\frac{\alpha-1}{\alpha+1}} \right)
\le \frac{\pi x}{L} \le \tan^{-1} \left(
\sqrt{\frac{\alpha-1}{\alpha+1}} \right)\right]. \label{fixedpt}
\end{eqnarray}
This range is a monotonically increasing function of $\alpha$, which
closes off near the transition line $\alpha=1$ and approaches $-L/4
\le x \le +L/4$ in the limit $\alpha \to \infty$. Clearly, for all
$x$-values outside this range, the corresponding OTOC will display
no non-trivial behaviour.

Unlike the scaling transformations, generated by the undeformed CFT
Hamiltonian on the plane, the M\"{o}bius transformation has fixed
points. The fixed points are by definition, points which do not
evolve under the M\"{o}bius transformation {\it i.e.} $z_{n} =z$. From its
definition, it follows that, at these points $\eta$ vanishes.
Interestingly, the two end points of the range $x_{fp\pm} = \pm
(L/\pi) \tan^{-1}[ \sqrt{(\alpha-1)/(\alpha+1)}]$ are actually the
fixed points of the M\"{o}bius flow. Thus, as shown in Fig.\
\ref{figd1} below, when the two operators ($V$ and $W$) are on the
same side of the fixed points, we see nontrivial $\eta(n)$ flow,
while when the two operators are on opposite side of the fixed
points, there is no such nontrivial behaviour.  Thus, unsurpisingly,
the existence of fixed points  seem to control the nature of the
OTOC behaviour in these systems.
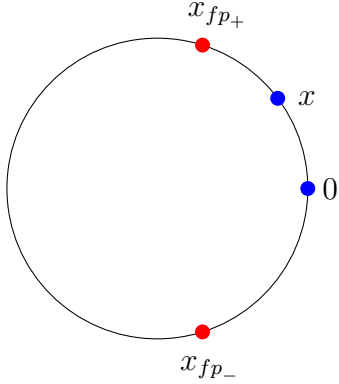
\begin{figure}
\centering
\begin{tikzpicture}
\draw (2,2) circle (2cm);
\draw (4.3,2) node {$0$};
\draw (3.99,3.2) node {$x$};
\draw (2.8,4.3) node {$x_{{fp}_{+}}$};
\draw (2.7,-0.4) node {$x_{{fp}_{-}}$};
\node[fill=blue,circle,inner sep=2pt,minimum size=1pt] at (4,2) {};
\node[fill=blue,circle,inner sep=2pt,minimum size=1pt] at (3.6,3.2) {};
\node[fill=red,circle,inner sep=2pt,minimum size=1pt] at (2.6,3.9079) {};
\node[fill=red,circle,inner sep=2pt,minimum size=1pt] at (2.6,0.0921) {};
\end{tikzpicture}
\caption{Schematic representation of fixed points on the circle. The
two red dots correspond to the two fixed points, while the blue
dots label the position of the two operators. For non trivial behaviour, both the operators have to be on
the same side of the fixed points.} \label{figd1}
\end{figure}

We now study the explicit form of the cross ratio $\eta$. This can
be done by substituting Eq.\ \ref{abcddef} in Eq.\
\ref{eta}, and setting $m=0$. A straightforward calculation yields
\begin{eqnarray}
\eta = \frac{2 \sin(\epsilon_{12}) \sin(\epsilon_{34}) \left( \alpha - 2 e^{\frac{2i\pi x}{L}} + \alpha e^{\frac{4 i\pi x}{L}}\right) }{ \left[\sqrt{\alpha+1} \left( 1 - e^{\frac{2i\pi x}{L}}\right) \cosh(n\theta + i \epsilon_{13}) + i \sqrt{\alpha-1} \left( 1 + e^{\frac{2i\pi x}{L}}\right) \sinh(n\theta + i \epsilon_{13}) \right] \left[ 1\to 2, 3\to 4\right]} \ . \label{crossratio}
\end{eqnarray}
To determine the large time behaviour, let us note that the
cross-ratio in (\ref{crossratio}), in the limit $n\theta \to
\infty$, behaves as: $\eta = \epsilon_{12} \epsilon_{34} A(x) e^{- 2
n \theta}$
\begin{eqnarray}
A(x) &=&  -\frac{4  \left(\alpha  \cos \left(\frac{2 \pi
x}{L}\right)-1\right)}{\left(\sqrt{\alpha -1} \cos \left(\frac{\pi
x}{L}\right)-\sqrt{\alpha +1} \sin \left(\frac{\pi
x}{L}\right)\right)^2} \ . \label{axeqn}
\end{eqnarray}
We note that $A(x)$ is purely negative for all $|\alpha|>1$.
\footnote{Note that this expression is valid only when $x$ is not
one of the fixed points. As mentioned previously at the fixed point,
$\eta$ is identically zero for all $n$.} The above can subsequently
be substituted to $F(\eta)$, when such a dynamical function is
known. Let us use the simplest example: large-$c$ CFT, with $h_w/c$
held small and fixed and $h_v$ fixed. In this case, using the
identity block dominance, we have (\cite{Fitzpatrick:2014vua},
\cite{Roberts:2014ifa})
\begin{eqnarray}
F(\eta ) = \left( \frac{1}{1 - \frac{24 i \pi h_w}{c \eta}} \right)^{2h_v} \ .
\end{eqnarray}
In the limit where $\frac{24 \pi h_w}{c \eta} \ll 1$, this becomes:
\begin{eqnarray}
F(t, x)  =  1 + \frac{48 i \pi h_w h_v }{c \epsilon_{12}
\epsilon_{34}} {\rm Exp}\left[\lambda_{\rm L} \left( t -
\frac{1}{\lambda_{\rm L}}\log(- A(x)) \right)\right]  + \cO(c^{-2}) \ ,
\label{otocgen}
\end{eqnarray}
 where $t = nT$ and the Lyapunov exponent
$\lambda_{\rm L}$ is given by $\lambda_{\rm L} = 2\theta/T$. We note that
this expression makes sense only if $c \eta \gg 24 \pi h_w$. It is
customary to think that this is guaranteed by large $c$; however,
this is not the case if $\alpha \gg 1$, {\it i.e.} near $\delta f
T/L \simeq -n$. For the rest of this section, we shall focus in the
regime where $\alpha$ is such that $c \eta \gg 24 \pi h_w$; we note
that this problem does not arise in the limit of high drive
frequency where $\alpha$ do not diverge for non-zero $\delta f$.

In this limit, one can define a position dependent butterfly
velocity $v_{\rm B}(x)$ and scrambling time $t^*(x)$ as follows.
 We first define
\begin{eqnarray}
\phi(x,t) \equiv t -\frac{1}{\lambda_{\rm L}}\log(-A(x))
-\frac{1}{\lambda_{\rm L}}\log(c), \label{phidef}
\end{eqnarray}
so that the butterfly velocity is defined as the velocity at which
$\phi(x,t)$ is an extremum {\it i.e.}~ $\frac{d\phi}{dt} =0.$ This reduces
to the usual definition when $\phi$ is linear in $x$, as is the case
for 2D thermal CFT at large $c$. The scrambling time is defined as
the time at which $(\phi(t^*, x)\sim 0).$ This reduces to what is
referred to as the ``relevant scrambling time " in the 2D CFT case
\cite{Roberts:2014ifa}. With these definitions, we get
\begin{eqnarray}
&& \quad v_{\rm B}(x) =  \lambda_{\rm L} \frac{A(x)}{A'(x)} = \frac{ \lambda_{\rm L} L}{2\pi} \frac{1}{\sqrt{\alpha^2 -1 }} \left( \alpha \cos\left(\frac{2\pi x}{L} \right) -1 \right) \ ,\\
&& t^*(x) = \frac{1}{\lambda_{\rm L}}  \left( \log c + \log (-A(x))  \right) \ .   \label{LyaButter}
\end{eqnarray}
We note that the conformal symmetry of the driven CFT forces a global drive protocol to explicitly
break spatial
translational invariance. Such a translational symmetry breaking
manifests itself as inhomogeneous structures in the correlation
functions. This is a generic feature in such driven systems. In the
context of OTOC, as we find here, this leads to a spatial dependence
of the butterfly velocity. In contrast, the scrambling time has weak
spatial dependence. This is due to the fact that $\log(- A(x))$ is
usually small compared to $\log c$ for a large-$c$ system leading to
a small spatial variation of the scrambling time.

\begin{figure}
\includegraphics[width=0.49\linewidth]{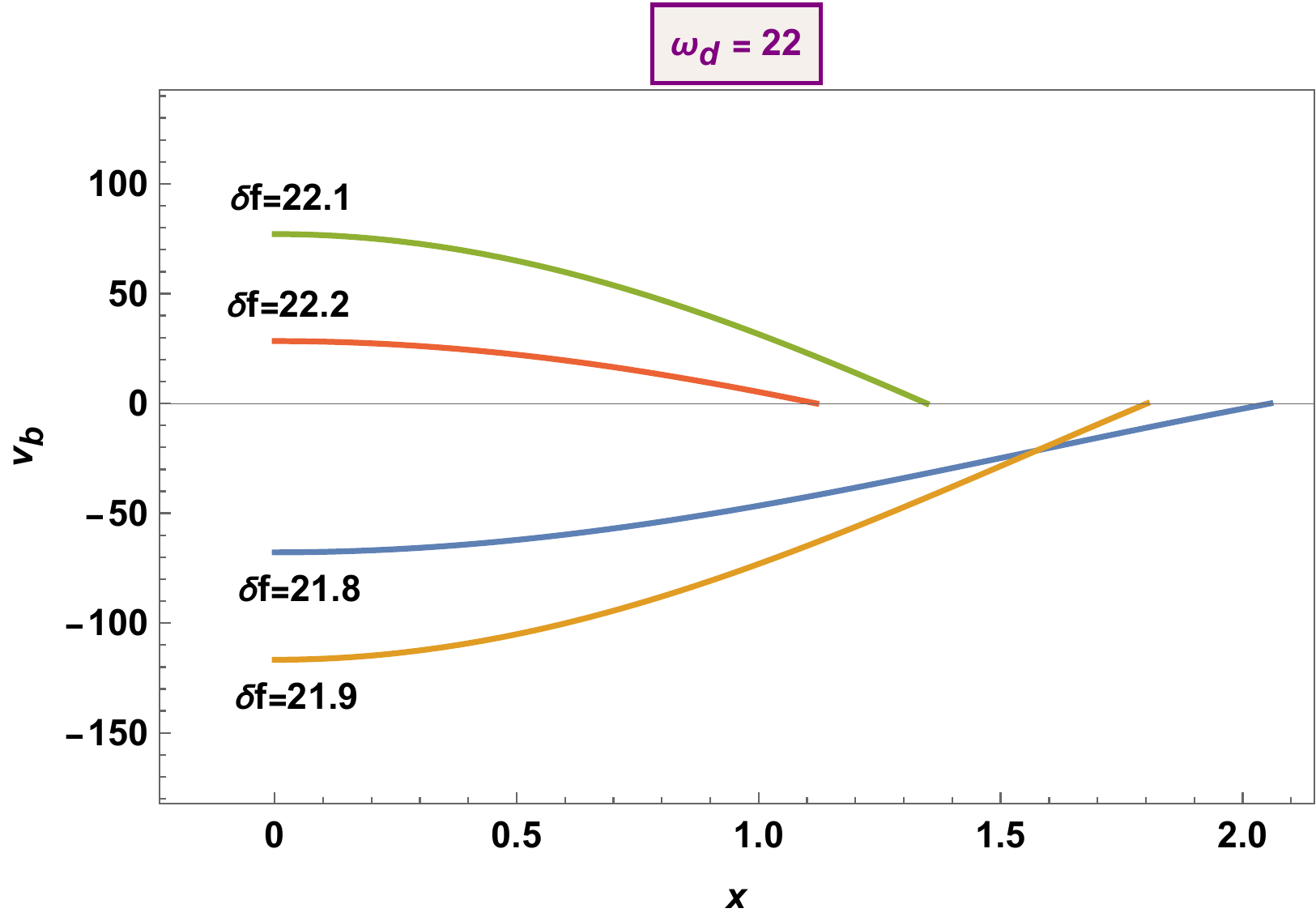}
\includegraphics[width=0.49\linewidth]{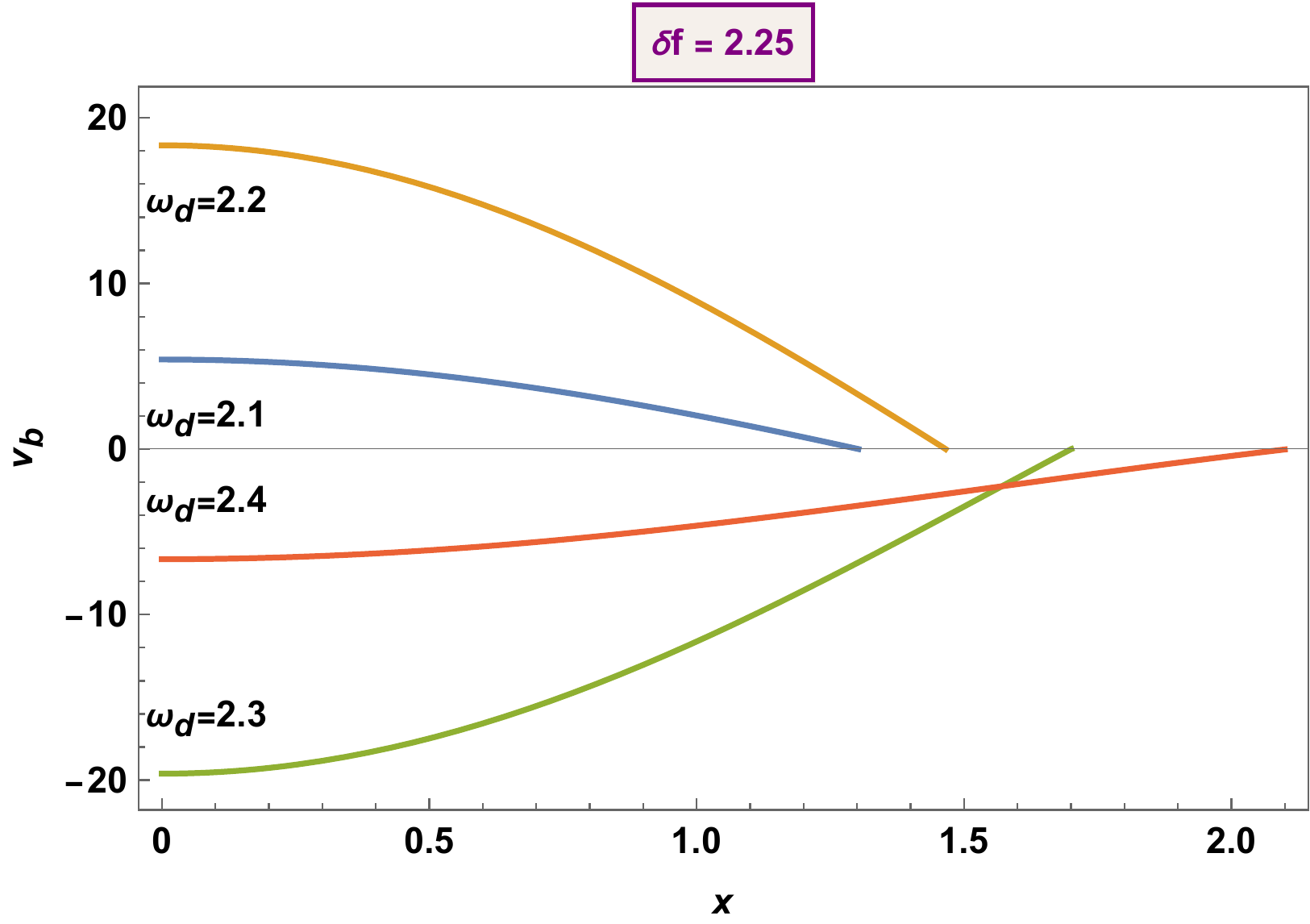}
\includegraphics[width=0.49\linewidth]{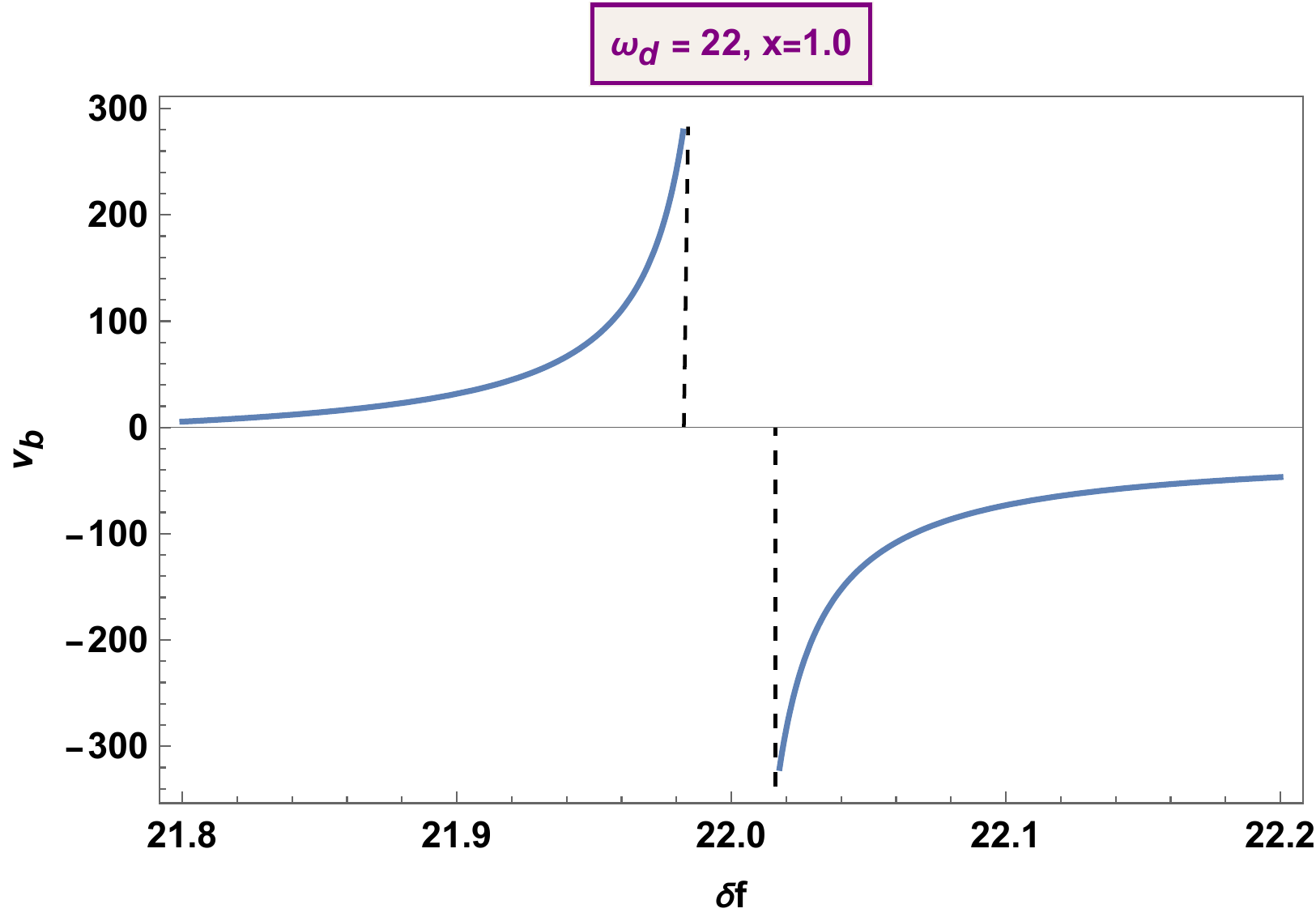}
\includegraphics[width=0.49\linewidth]{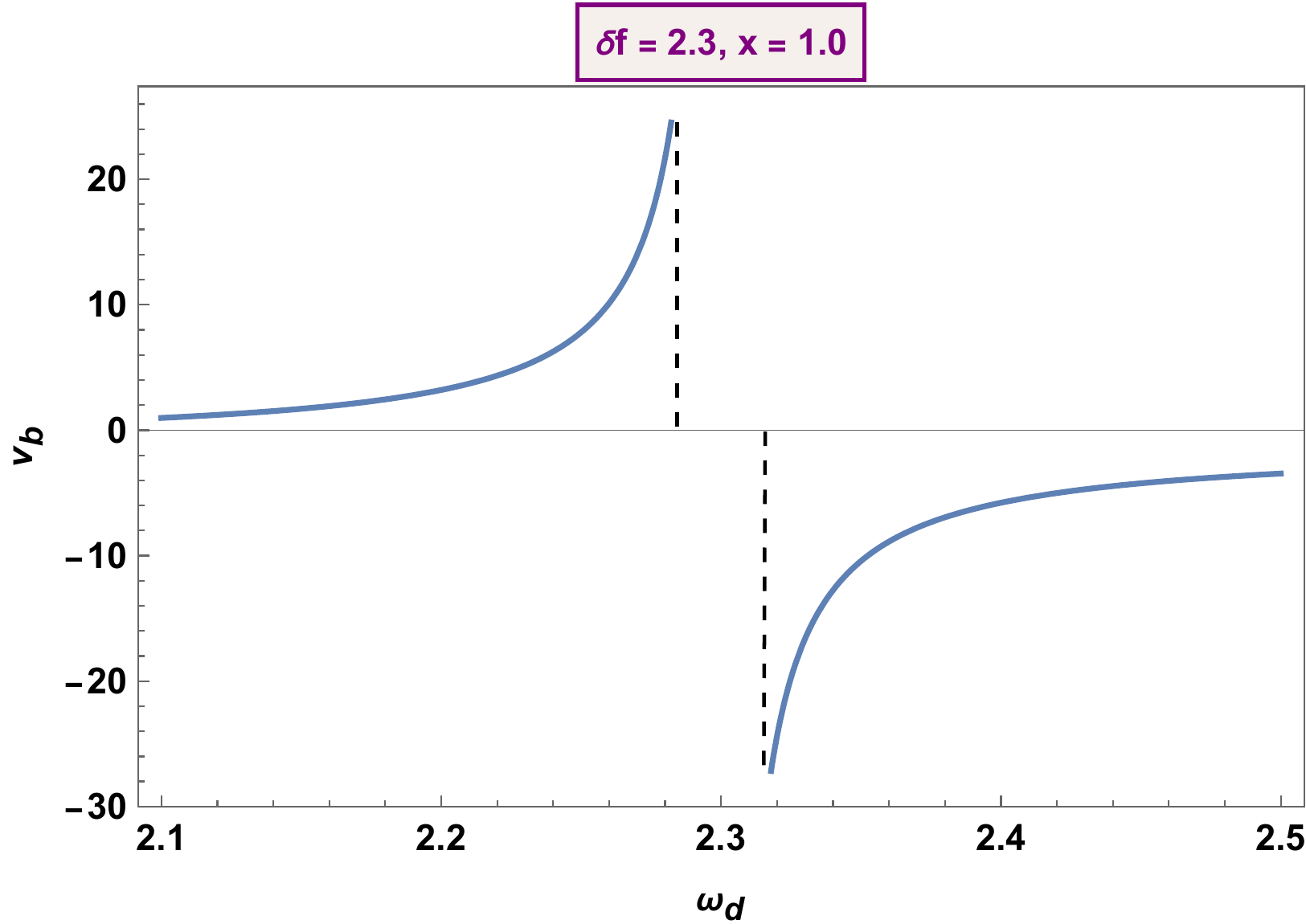}
\caption{Top left panel: Spatial variation of the butterfly velocity
$v_{\rm B}(x)$ as a function of $x$ and several $\delta f$ for
$\omega_D=2\pi/T=22$. The end points of the curves indicate the
range of $x$ where $v_{\rm B}(x)$ can be defined. Top right panel:
Same as the top left panel but for different $\omega_D$ and $\delta
f=2.25$. Bottom left panel: Plot of $v_{\rm B}$ at $x=1$ for as a
function of $\delta f$ for $\omega_D$. The dashed lines are
schematic guide to the eye representing lines near $\omega_D=\delta f$
where $c \eta \simeq 24 \pi h_w$. Bottom right panel: A similar plot
as the left panel but as a function of $\omega_D$ and for $\delta
f=2.3$. All distances are scaled in units of $\frac{L}{2\pi}$ and
all energies are scaled in units of $\frac{2\pi}{L}$. The drive
amplitude is set to $f_1=10$ for  all figures.
 } \label{fig1}
\end{figure}

The spatial variation of the butterfly velocity is shown in the top
panels of Fig.\ \ref{fig1}. We find that depending on the drive
frequency (for a fixed $\delta f$) or $\delta f$ (for a fixed drive
frequency), $v_{\rm B}(x)$ may either increase or decrease as a
function of $x$. The end points of these curves indicate the fixed
points defined in Eq.\ \ref{fixedpt}. The plot of
$v_{\rm B}$ for a fixed $x$ is shown in the bottom panels as a
function of $\delta f$ (left) and $\omega_D$ (right). The horizontal
dashed line indicate the schematic marks for $c \eta \simeq 24 \pi
h_w$; between these lines, the values of $|\alpha|$ are large enough
to invalidate the expansion of $F(x)$. We find that the sign of
$\alpha$ and hence $v_{\rm B}(x)$ (Eq.\ \ref{LyaButter}) changes
across these regions; thus the direction of the butterfly velocity
at a given spatial point can be tuned by tuning either $\omega_D$ or
$\delta f$. To the best of our knowledge, this feature has no
analogue in homogeneous CFTs.

%%%%%%%%%%%%%%%%%%%%%%%%%%%%
\subsubsection{$4$-point OTOC for generic configurations.}
%%%%%%%%%%%%%%%%%%%%%%%%%%%%

The initial positions of the operators taken in the above
discussion, ($x$ and $0$) are not the most general. In the
undeformed CFT case, due to translational invariance, one can
without loss of generality choose one operator to be at the origin.
However, the drive is not spatially homogeneous and so the most
general configuration would be when the two operators are placed at
two arbitrary points ($x_1$ and $x_2$), with the corresponding configurations on the plane.
\begin{eqnarray}
z_1 = e^{\frac{2\pi ix_1}{L}} = z_2 \ , \quad z_3 = e^{\frac{2\pi i x_2}{L}} = z_4 \ .
\end{eqnarray}
Proceeding with the definition in (\ref{eta}), we obtain the
following expression for the cross-ratio $\eta$:
\begin{eqnarray}
&& \eta = \frac{ \epsilon_{12} \epsilon_{34} \left(\alpha - 2 D_1 + \alpha D_1^2 \right) \left( D_1 \to D_2\right) }{\left[ \left( D_1 - D_2 \right) \sqrt{\alpha^2-1} \cosh\left( n \theta + i \epsilon_{13}\right) + i \left( \alpha+ \alpha D_1 D_2 -D_1 - D_2 \right) \sinh\left( n \theta + i \epsilon_{13} \right) \right] \left[ \epsilon_{13} \to \epsilon_{24}\right] } \ , \nonumber\\
\end{eqnarray}
where $ D_{1,2} = \exp[2\pi i x_{1,2}/L]$. The corresponding
kinematics is no longer a function of $(x_1 - x_2)$, as is also
obvious from the expression of the cross-ratio above. As before, the
$\eta$ would vanish if the initial position of either of the
operators is at the fixed point of the M\"{o}bius flow. Also, like
before, $\eta$ crosses the real axis at the point when $(z_{1n}
=z_3)$. The corresponding condition is
\begin{eqnarray}
\tanh\left( n \theta \right)  = i \frac{ \left( D_1 - D_2\right)  \sqrt{\alpha^2 -1} }{\alpha + \alpha D_1 D_2 - D_1 - D_2} \ . \label{realx1x2}
\end{eqnarray}
The corresponding condition for $\bar{\eta}$ is :
\begin{eqnarray}
\tanh\left( n \theta \right)  = -i \frac{ \left( D_1 - D_2\right)  \sqrt{\alpha^2 -1} }{\alpha + \alpha D_1 D_2 - D_1 - D_2} \ . \label{barrealx1x2}
\end{eqnarray}

It is easy to check that setting $D_1=0$ in (\ref{realx1x2} and
\ref{barrealx1x2}) yields (\ref{cond1} and \ref{cond2}). It is also
straightforward to check that, when (\ref{realx1x2}) is satisfied,
one obtains: $\eta = (\epsilon_{12}\epsilon_{34})/(\epsilon_{13}
\epsilon_{24})$. As before, one can solve (\ref{realx1x2}) to find a
range of values for $x_1$ and $x_2$.   For this purpose, it's
convenient to define the variables $X =(x_1 + x_2) /2$ and $x
=(x_1-x_2)/2$. The range of these variables is $(-X\leq x\leq X)$
and $ 0\leq X <2\pi$. In terms of these variables, Eqs.\
\ref{realx1x2} and \ref{barrealx1x2} maybe rewritten as:
\begin{equation}\label{z1n=z3}
\tanh(n\theta) = \pm \frac{\sin(\frac{2\pi x}{L})\sqrt{\alpha^2-1}}{\cos(\frac{2\pi x}{L}) - \alpha \cos(\frac{2\pi X}{L})}\; ,
\end{equation}
where the $+(-)$ signs correspond to the conditions from $\eta$ and
$\bar{\eta}$ respectively. The range of the $x_1$ and $x_2$ for
which this is valid is given by:
\begin{equation}
1 \leq \frac{\sin(\frac{2\pi x}{L})\sqrt{\alpha^2-1}}{\cos(\frac{2\pi x}{L}) - \alpha \cos(\frac{2\pi X}{L})} \leq -1\; .
\end{equation}

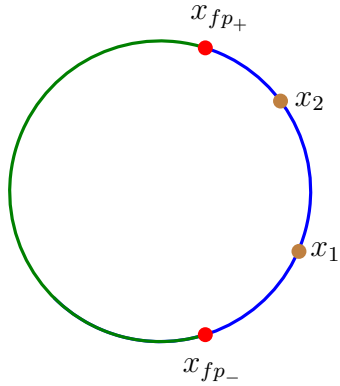
\begin{figure}\label{4}
    \centering
    \begin{tikzpicture}
        \draw (2,2) circle (2cm);
        \draw (4.2,1.2) node {$x_1$};
        \draw (3.99,3.2) node {$x_2$};
        \draw (2.8,4.3) node {$x_{{fp}_{+}}$};
        \draw (2.7,-0.4) node {$x_{{fp}_{-}}$};
        \draw [very thick,blue] (2,2) ++(75:2.0) arc[start angle=75,delta angle=-210,radius=2];
        \draw [very thick,darkgreen] (2,2) ++(75:2.0) arc[start angle=75,delta angle=210,radius=2];
        \node[fill=brown,circle,inner sep=2pt,minimum size=1pt] at (3.845,1.2) {};
        \node[fill=brown,circle,inner sep=2pt,minimum size=1pt] at (3.6,3.2) {};
        \node[fill=red,circle,inner sep=2pt,minimum size=1pt] at (2.6,3.9079) {};
        \node[fill=red,circle,inner sep=2pt,minimum size=1pt] at (2.6,0.0921) {};
    \end{tikzpicture}
    \caption{Schematic representation of two regions marked in the region plot as A and B. The green part of the circle represents B while the blue part represents A. Fixed points on the circle are marked with
        two red dots. The brown dots label the positions of the two operators at $x_1$ and $x_2$. For non trivial behaviour, both the operators have to be on
        the same side of the fixed points (Fig.\ \ref{figd4}). In this figure both points are in region
        A} \label{figd2}
\end{figure}

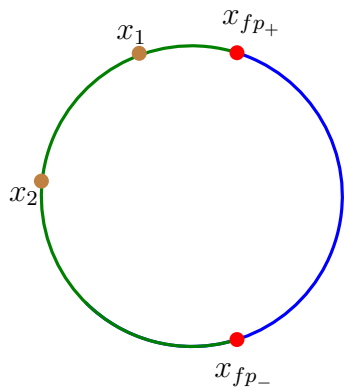
\begin{figure}\label{5}
    \centering
    \begin{tikzpicture}
        \draw (2,2) circle (2cm);
        \draw (1.2,4.15) node {$x_1$};
        \draw (-0.23,2.0) node {$x_2$};
        \draw (2.8,4.3) node {$x_{{fp}_{+}}$};
        \draw (2.7,-0.4) node {$x_{{fp}_{-}}$};
        \draw [very thick,blue] (2,2) ++(75:2.0) arc[start angle=75,delta angle=-210,radius=2];
        \draw [very thick,darkgreen] (2,2) ++(75:2.0) arc[start angle=75,delta angle=210,radius=2];
        \node[fill=brown,circle,inner sep=2pt,minimum size=1pt] at (1.3,3.9) {};
        \node[fill=brown,circle,inner sep=2pt,minimum size=1pt] at (0.0,2.2) {};
        \node[fill=red,circle,inner sep=2pt,minimum size=1pt] at (2.6,3.9079) {};
        \node[fill=red,circle,inner sep=2pt,minimum size=1pt] at (2.6,0.0921) {};
    \end{tikzpicture}
    \caption{Same as Fig.\ \ref{figd2} but with both points in region B. This corresponds to the shaded center square of
    Fig.\ \ref{figd4}.} \label{figd3}
\end{figure}

\begin{figure}\label{7}
    \centering
    \includegraphics[width=0.49\linewidth]{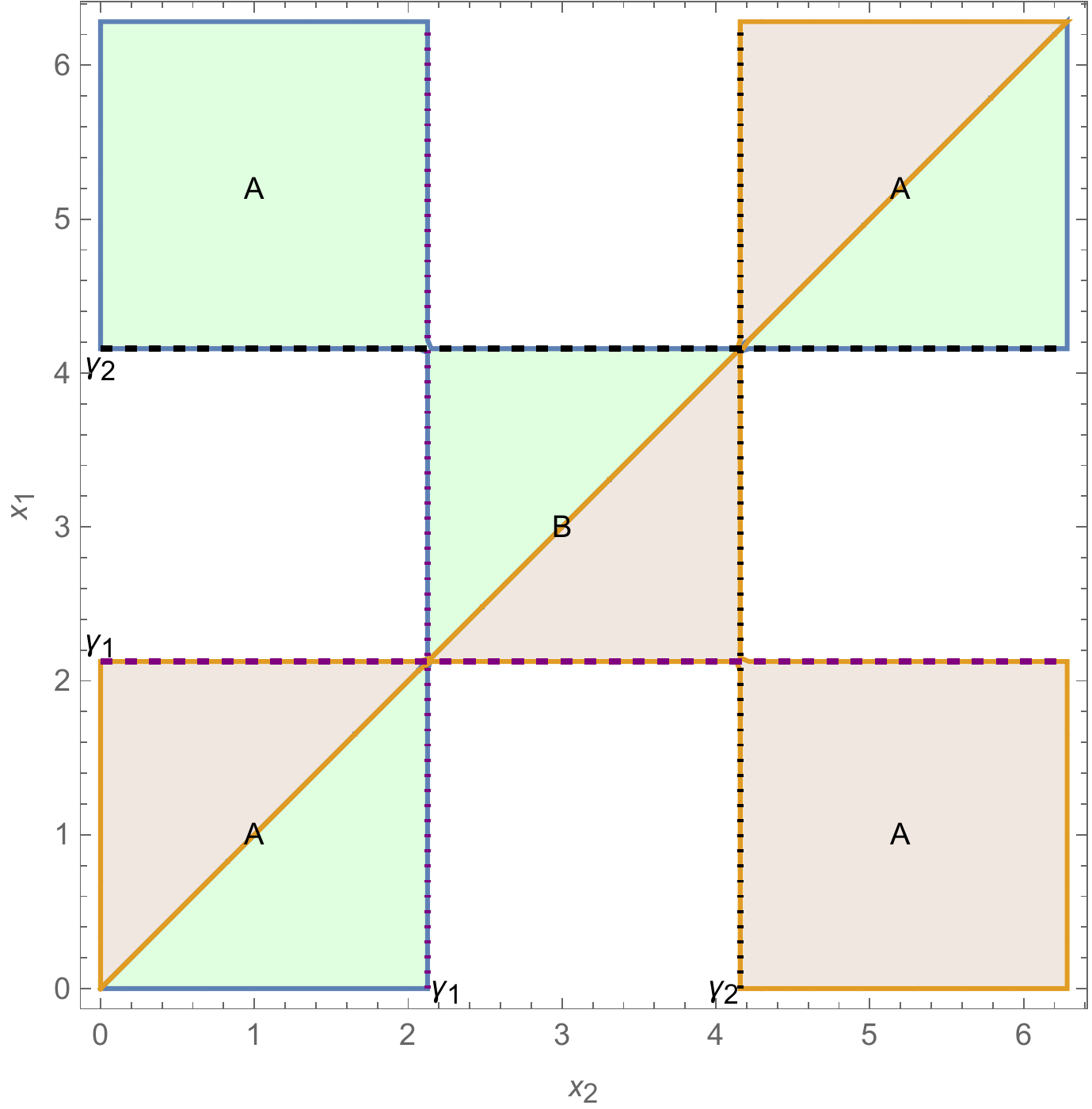}

    \caption{ Plot showing the allowed regions in $x_1$ and $x_2$. The green regions indicate the
    allowed ranges for $\eta$ while the brown regions indicate that for $\bar{\eta}$. $\gamma_1$ and $\gamma_2$ label the fixed points of the flow (These are the same points that are labelled as $x_{fp\pm}$ on the circle in Figs.\ \ref{figd1}, \ref{figd2} and \ref{figd3}). All distances are scaled in units of $\frac{L}{2\pi}$.
    } \label{figd4}
\end{figure}

The explicit solution space of the above inequality is shown for a particular value of $\alpha$ in Fig.\
\ref{figd4}. Here $\gamma_1$ and $\gamma_2$ are the two fixed
points. The solutions are given by the shaded region in Fig.\
\ref{figd4}. This corresponds to both points being on the same side
of the fixed points. This is schematically represented in Figs.\ \ref{figd2} and \ref{figd3}. Indeed this is as it should be, since
$z_1(n)=z_3$ only if initially $z_1$ and $z_3$ were placed on the
same side of the fixed points, else $z_1$ would end up at the fixed
point and Eq.\ \ref{z1n=z3} could not have been satisfied.

One can also obtain the large time behaviour, we take $n \to
\infty$, which yields: $\eta = \epsilon_{12} \epsilon_{34} e^{- 2 n
\theta} A(x_1, x_2)$
\begin{eqnarray}
\textrm{where,}\; \;  A(x_1,x_2) &=& 4\frac{\left( \alpha - 2 D_1 + \alpha D_1^2\right)
\left( D_1 \to D_2 \right)}{\left[ \left( D_1 - D_2 \right) \sqrt{\alpha^2 -1}
+ i \left(\alpha + \alpha D_1 D_2 - D_1 - D_2 \right)  \right]^2} \label{ax1x2eq} \\
&=& -4\frac{(\alpha\cos(\frac{2\pi x_1}{L})-1)(\alpha\cos(\frac{2\pi x_2}{L})-1)}
{(\sqrt{\alpha -1}\cos(\frac{\pi x_2}{L}) -\sqrt{\alpha+1}\sin(\frac{\pi x_2}{L}))^2
(\sqrt{\alpha_1}\cos(\frac{\pi x_1}{l})+\sqrt{\alpha +1}\sin(\frac{\pi x_1}{L}))^2}\nonumber
\end{eqnarray}

With this information, one can again find the OTOC for a large $c$
CFT. The Lyapunov exponent is again given by (\ref{LyaButter}).
There is however an ambiguity in the definition of the butterfly
velocity, since now there are two positions $x_1$ and $x_2$.
Intuitively, it should be a measure of how fast information spreads
between the two points. There are two possible definitions,
consistent with this intuition. One way to define it is as-
$v^{-}_{\rm{B}} = 2\frac{dx_-}{dt}|_{x_+=\rm{constant}}$. The second
possible definition would be as:
$v_{\rm{B}}=\frac{dx_{2}}{dt}|_{x_1=\rm{constant} }$. This second
definition also has the advantage that it reduces to the definition
of the previous section.

\begin{eqnarray}
&&\frac{d}{dt} \left(  t - \frac{\log A(x_1, x_2)}
{\lambda_{\rm L}} \right)\Big |_{x_1 =\rm{constant}}  = 0 \quad \implies \quad 1
- \frac{1}{\lambda_{\rm L}} \left( \frac{\partial_2 A}{A} v_{\rm B}\right) = 0 \nonumber \\
&& \quad \implies v_{B} =\lambda_{\rm L} \left( \frac{\partial_2 A}{A}\right)^{-1} =\lambda_{\rm L} \frac{L}{2\pi}\frac{\left(\alpha\cos(\frac{2\pi x_2}{L})-1\right)}{\sqrt{\alpha^2-1}}
\end{eqnarray}

Remarkably, with the second definition, the butterfly velocity turns
out to be independent of the position $x_1$ and exactly matches with
the previous case $x_1=0$. This is a direct
consequence of the fact $A(x_1,x_2)$ can be written as products of
functions of $x_1$ and $x_2$ at late times as can be seen from Eq.\
\ref{ax1x2eq}.

%%%%%%%%%%%%%%%%%%%%%%%%%%%%%%%
\subsubsection{On the Transition Line: $\alpha =1$}
%%%%%%%%%%%%%%%%%%%%%%%%%%%%%%%

The analysis above assumes $|\alpha| > 1$. On the transition line, {\it i.e.} when $\alpha =1$, we can write\cite{Das:2021gts}:
\begin{eqnarray}
\frac{1}{p} \approx \alpha \ , \quad p \approx \frac{s}{T} \ ,
\end{eqnarray}
and subsequently, the coefficients are:
\begin{eqnarray}
a_n = 1 - i p n T = 1 - i n s \ , \quad d_n = 1 + i n s \ , \quad b_n = - i n s = - c_n \ .
\end{eqnarray}
With these, one now obtains:
\begin{eqnarray}
\eta = \frac{4 s^2 e^{\frac{2\pi i x}{L}} \epsilon_{12}\epsilon_{34}}{\left[ 1 - 2 i n s + 2 s \epsilon_{13} - e^{\frac{2\pi i x}{L}} \left(1 + 2 i n s - 2 s \epsilon_{13} \right) \right] \left[ 1\to 2, 3\to 4\right]} \ . \label{eta0}
\end{eqnarray}
Proceeding as before, setting the $z_{1n}= z_{3}$, we get:
\begin{eqnarray}
\cos \left(\frac{2\pi i x}{L} \right)  =  \left( \frac{1 - 4 n^2 s^2 }{1+ 4 n^2 s^2 }\right) \ ,  \label{condx0}
\end{eqnarray}
Using (\ref{condx0}) in (\ref{eta0}), we again obtain:
\begin{eqnarray}
\eta = \frac{\epsilon_{12} \epsilon_{34}}{\epsilon_{13}\epsilon_{24}} \ .
\end{eqnarray}
Evidently, in the limit $n\to 0$, $\eta \sim \cO(\epsilon^2)$ and in
the limit of large $n$, $\eta \sim
A(x)n^{-2}\epsilon_{12}\epsilon_{34}$.\footnote{Note that, by
definition $n\in {\mathbb Z}_+ \cup \{0\}$. However, $nT$, where $T$
is the period of the drive, is a real-valued parameter. Thus, all
smooth limits {\it e.g.}~$n\to 0$ or large $n$ is to be understood
as $nT \to 0$ and $nT\to \infty$, respectively. } Here $A(x) =
-\frac{e^{\frac{2\pi i x}{L}}}{(1+e^{\frac{2\pi i x}{L}})^{2}}$.
Hence the OTOC in this phase transition regime shows power law
growth of $\mathcal{O}(n^{2})$. The absence of Lyapunov growth is
also a feature of non-heating phase ($\alpha^{2}<1$), where one
could obtain oscillatory behavior of OTOC in a similar manner by
using appropriate $a_{n},b_{n},c_{n},d_{n}$ as discussed in section
\ref{sec cont}.

%%%%%%%%%%%%%%%%%%%%%%%
\subsection{4pt OTOC in the Discrete Driven CFTs}\label{discotoc}
%%%%%%%%%%%%%%%%%%%%%%%

\begin{figure}\label{8}
    \includegraphics[width=0.40\linewidth]{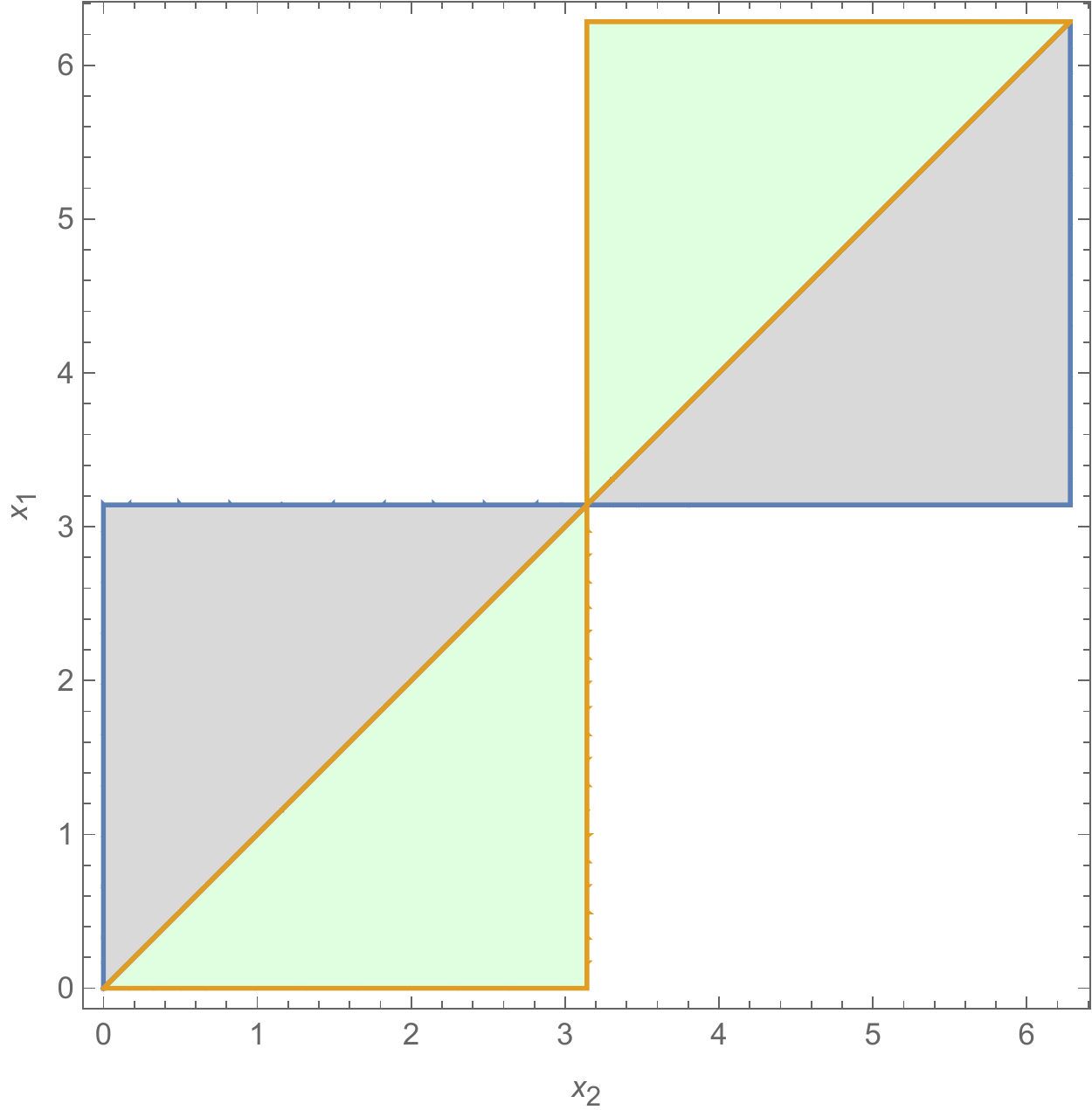}
    \includegraphics[width=0.60\linewidth]{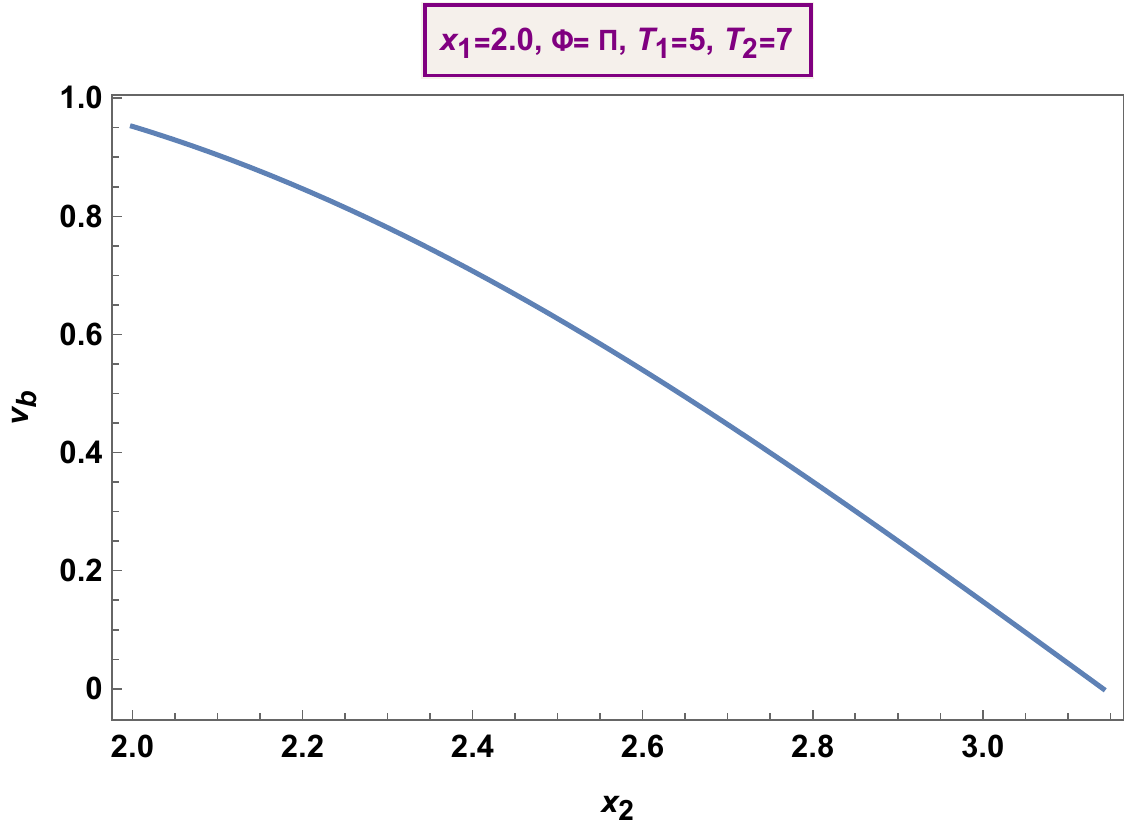}
\caption{Left: Plot showing the allowed regions in $x_1$ and $x_2$
in discrete drive case. The green regions indicate the allowed
ranges for $\eta$ while the grey regions indicate that for $\bar{\eta}$.
Right: Change in $v_{\rm B}$ with $x_2$ when $x_1$ is kept fixed.
All distances are scaled in units of $\frac{L}{2\pi}$.} \label{fig4}
\end{figure}

We now consider the case of the discrete drive similar to the ones
considered in \cite{Wen:2020wee}. The corresponding Hamiltonian is
given in Eq.\ \ref{discrete}. Let us begin with the initial operator
locations: $z_1 = z_2 = e^{\frac{2\pi i x_1}{L}}\equiv D_1$ and $z_3
= z_4 = e^{\frac{2\pi ix_2}{L}}\equiv D_2$. The corresponding
cross-ratio is given by
\begin{eqnarray}
\eta = - \frac{\left(D_1^2 -1 \right) \left(D_1^2 -1 \right) \sin\left(2 \phi \epsilon_{12}\right) \sin\left( 2\phi \epsilon_{34}\right)}{\left[ \left( D_1D_2 - 1\right) \sinh\left(2\phi(n + i \epsilon_{13}  \right) + \left(D_1 - D_2 \right) \cosh\left( 2\phi(n + i \epsilon_{13} \right)   \right] \left[ \epsilon_{13} \to \epsilon_{24}\right]} \ .
\end{eqnarray}
Interestingly, the cross ratio in this case vanishes if we take
either $x_1=0$ or $x_2=0$. This is because $x=0$ is a fixed point of
the flow. This situation is schematically shown in the left panel of
Fig.\ \ref{fig4}.

Repeating the analysis from the previous cases, we find that the
condition for $z_{1n} =z_{3}$ gives us:
\begin{eqnarray}
\tanh\left( 2 n \phi\right)  = \frac{D_1 - D_2}{1 - D_1 D_2} =
-\frac{\sin(\frac{\pi(x_1-x_2)}{L})}{\sin(\frac{\pi (x_1+x_2)}{L})}
\ . \label{eqrela}
\end{eqnarray}

As in the continuous drive example, for Eq.\ \ref{eqrela} to have
any solution, $x_1$ and $x_2$ have to be on the same side of the
fixed points. Once again, as was the case previously, when this
condition is met we see that the cross ratio just crosses the real
axis at the point $\eta =
\frac{\epsilon_{12}\epsilon_{34}}{\epsilon_{13}\epsilon_{24}}$.
Thus, $\eta$ crosses the branch cut for OTOCs. Including the
condition from $\bar{\eta}$, as before, we get the full range of
values of $x_i$ for which  we have non-trivial OTOCs. This is given
by:
\begin{equation}
-1 \leq -\frac{\sin(\frac{\pi(x_1-x_2)}{L})}{\sin(\frac{\pi (x_1+x_2)}{L})} \leq 1 \ .
\end{equation}

In the large $n$ limit,
\begin{eqnarray}
&& \eta = \epsilon_{12}\epsilon_{34} e^{- 4 n \phi} A(x_1, x_2) \ , \\
\textrm{where:}\; && A(x_1,x_2) = - 4 \frac{\left(D_1^2 -1 \right)\left(D_2^2 -1 \right)\left( 2 \phi \right)^2}{\left[ \left(D_1 D_2 -1 \right) + \left( D_1 - D_2 \right) \right]^2 }\, \\
&&  = -16\;\phi^2\frac{\tan(\frac{\pi x_2}{L})}{\tan(\frac{\pi x_1}{L})} \ .
\end{eqnarray}
The large $c$ CFT analysis is the same, except for a change in the
form of  $A(x_1,x_2)$. In this case we get $\lambda_{\rm L}
=\frac{4\phi}{T_1 +T_2}$ and (using the second definition for the
butterfly velocity ($v_{\rm B}=\frac{\lambda_{\rm
L}}{\frac{\partial\ln(A)}{\partial x_2}}$))
\begin{equation}
v_{\rm B}
= \lambda_{\rm L}\frac{L}{2\pi}\sin \left(\frac{2\pi x_2}{L}\right )  \ .
\end{equation}
A plot of the butterfly velocity for the discrete drive protocol is
shown in the right panel of Fig.\ \ref{fig4}; as before, this is
independent of $x_1$.

%%%%%%%%%%%%%%%%%%%
\numberwithin{equation}{section}
\section{A Holographic Perspective}
%%%%%%%%%%%%%%%%%%%

Let us now offer some comments regarding how such driven systems may
be realized in the Holographic context. Our goal here is to offer
comments that can bridge between the existing literature. First,
note that, given an initial thermal state in the dual CFT and a
particular drive protocol in any dimensions, the Holographic
description is a well-posed problem in an asymptotically AdS-space.
For example, in \cite{Rangamani:2015sha, Biasi:2017kkn, Biasi:2019eap} a deformation of the
following kind has been considered:
\begin{eqnarray}
S = S_{J=0} + \int d^dx \sqrt{-\gamma} J(x) {\cal O}(x) \ ,
\end{eqnarray}
where ${\cal O}$ is a relevant operator in the CFT. The coupling
$J(x)$ can be promoted to a function of time which defines the drive
protocol. For small amplitudes of the drive, the physics is
dissipation-dominated, since the system can dissipate the
drive-energy into its thermal bath. At large amplitudes, however,
this is not true and one obtains a rich phase structure.
Correspondingly, low-point correlation functions were also explored
in \cite{Rangamani:2015sha, Biasi:2017kkn, Biasi:2019eap}.

We can contrast this framework with what we have explored here. In
our framework, which is based on two-dimensional CFTs, the state is
taken to be a vacuum state and only the operators evolve. In doing
so, even two-point functions display a rich phase structure: for
example, the heating and the non-heating phases and the breaking of
spatial homogeneity. The dual picture in the three-dimensional
AdS-geometry is currently unknown. In view of the results in
\cite{Rangamani:2015sha, Biasi:2017kkn, Biasi:2019eap}, it seems that some salient features of the
driven CFT systems are currently unexplored in the dual
gravitational system.

One subtlety lies in formulating an analogous question in the
AdS-geometry within a Heisenberg picture. Typically, a classical
field in the bulk geometry back-reacts on the geometry and as a
result the geometry also becomes dynamical. Typically, given a
boundary scalar deformation, one expects that the gravitational
back-reaction becomes stronger as one approaches towards the IR,
which results in the formation of a black hole in the geometry.
Thus, intuitively, a heating phase is expected for an energy
injection at the conformal boundary of the AdS-geometry. However,
both low ({\it e.g.}~$2$-point) and higher-point functions ({\it
e.g.}~$4$-point OTOCs) in the dual CFT clearly demonstrate the
existence of a heating phase and a non-heating phase, by tuning
parameters of the drive protocol, which in our case is encoded in
$\alpha$. At least in three bulk dimensions, it is expected that a
bulk realization of this physics exists based on the well-known
Brown-Henneaux transformations.

While, generically, a black hole state is expected in the
Holographic context, it is less intuitive how one may obtain an
oscillatory correlation function. Note that, as far as an
oscillatory behaviour is concerned, thin time-like shells can
undergo an oscillatory motion in AdS, which yields a corresponding
behaviour in real-time for two point
functions\cite{Banerjee:2017lhg}. It was further noted that such
oscillating configurations rely on when the perfect fluid on the
shell is described by a polytropic equation of state. This is
perfectly consistent with weak energy condition in the bulk
geometry. It is therefore conceivable that by suitably adjusting the
equation of state of the fluid on the collapsing shell, one
interpolates between a heating phase and a non-heating phase.
Introducing an inhomogeneity in the system can further be modelled
by injecting the thin shell locally at the conformal boundary.

Going back to the issue of operator evolution in the Heisenberg
picture, a reasonable starting point is to construct local particle
states in AdS, which can subsequently be subject to an appropriate
evolution. In \cite{Berenstein:2019tcs}, both local states and
scattering states in global AdS were constructed for (regularized)
operators with large conformal dimensions. In particular,
\cite{Berenstein:2019tcs} considers operators of the form $\int
d\Omega {\cal O}_{\Delta,\epsilon} |0\rangle$, where $d\Omega$ is
the solid angle corresponding to an $S^{(d-1)}$ (on which the CFT is
defined) and ${\cal O}_{\Delta, \epsilon}(t) = e^{-\epsilon H} {\cal
O}_{\Delta}(t) e^{\epsilon H}$ defines a UV-regulator $\epsilon$. As
argued in \cite{Berenstein:2019tcs}, this corresponds to single
particles localized within subAdS scales of size $\sim
1/\sqrt{\Delta}$, where $\Delta$ is the conformal dimension of
${\cal O}_{\Delta}$.

Insertion of the operator $\int d\Omega {\cal O}_{\Delta,\epsilon}
|0\rangle$ at an initial time slice $t=0$ provides the boundary
condition for the bulk particle, which is at rest. The initial
position of the particle is set by the CFT UV-regulator $\epsilon$.
Since the inserted operator preserves a rotational symmetry along
the CFT directions, the dual single particle can evolve only by
traversing a radially inward geodesic. Correspondingly, time
evolution of this operator is dual to the bulk infalling particle in
this picture. This is particularly instructive, since OTOCs in the
dual CFT are eventually obtained in the geometric picture from
amplitudes of elastic $2-2$ scattering in the eikonal approximation.
While this is a promising scenario in which a Holographic dual
description of driven CFTs can be realized, it is necessarily
defined only within an asymptotically global AdS geometry. {\it A
priori}, the infalling particle should encode the drive protocol and
therefore potentially it offers a richer structure than what one
might obtain from a $2-2$ scattering near a black hole. The latter
is universally dictated by the large blue shift near the event
horizon and one obtains a universal answer determined by the surface
gravity. It is intriguing to note that there is no known geometric
dual (pure state, {\it e.g.}~a specific microstate geometry) in
which OTOCs can show oscillatory behaviour.\footnote{Note that even
extremal black hole background yield an exponentially growing
behaviour, see {\it e.g.}~\cite{Banerjee:2019vff, Craps:2020ahu}.}
Power-law behaviour --- this is what appears at the phase boundary
for $\alpha^2=1$ --- can be reproduced by a thermal AdS background.
In this respect, the driven system exhibits rather unique features.
It will be very interesting to better understand these aspects from
a Holographic perspective.

\numberwithin{equation}{section}
\section{Discussion}\label{discuss}

In this article, we have studied four point out of time ordered
correlation functions in  large $c$ driven CFTs. We find that these
functions show qualitatively different behaviour in the  {\it
heating} and {\it non-heating} phases. In the heating phase, the
OTOCs  show an exponential behaviour for times sufficiently smaller
than a suitably defined ``scrambling time", while in the non-heating
phase we see a oscillatory  behaviour. Both the ``scrambling time"
and a ``butterfly velocity" in our case are position dependent.
Moreover, unlike  the undeformed CFT Hamiltonian, the exponential
growth of the OTOC in the heating phase happens only when the
operators are placed within a special range of spatial coordinate
values determined by the fixed points of the M\" {o}bius flow. The
existence of this spatial profile in the OTOC and the associated
spatial dependence of the butterfly velocity are a consequence of
the spatial inhomogeneity introduced by the drive.

The Lyapunov exponent that we obtain from our analysis for the large
$c$ CFT matches exactly with
with the group theoretic Lyapunov exponent obtained
in\cite{Wen:2020wee} when applied for the continuous drive case, and
differs by a factor of 4 from the discrete drive case. Of course
these two notions are very different. The Lyapunov exponent obtained
in this note is intrinsically quantum mechanical, as it is defined
through a four-point function, while the group-theoretical Lyapunov
may be interpreted as a classical notion measuring the sensitivity
of the coordinate $z_f$ of a
quasi-particle to its initial position $z_i$ on the complex plane after $n$ iterations of
the M\"{o}bius evolution. Therefore a priori there is no reason to
expect them to match. Indeed the matching seems to be primarily a
consequence of dimensional analysis and the fact that $n$ always
enters into all expressions as $n\theta$.

The existence of novel phases like the heating and non-heating
phases is universal and exists for all Floquet CFTs
\cite{Han:2020kwp, Wen:2020wee}. However, as we show in the
appendix, for driven Ising CFT, the OTOC shows no exponential
behaviour at large values of the stroboscopic time, even in the
heating phase. We expect this to hold for other integrable CFTs as
well. For example, one could explicitly check that in driven minimal
models, at very late time the OTOC differs from TOC by a
constant(which is completely determined by modular S-matrix) upto an
additive exponential decay term \footnote{This is in the same spirit
of computing thermal OTOC in minimal models
\cite{Gu:2016hoy}-\cite{Fan:2018ddo}. The late time behavior of
thermal cross ratio is very similar to driven cfts except the
existence of novel spatial profile. Thus the analytic continuation
of Euclidean four point functions provides a leading constant OTOC
in the heating phase of driven cfts which is similar to thermal
OTOC.}. Thus our analysis shows that the OTO correlation function
continues to be a good diagnostic of chaos, demarcating driven
chaotic CFT Hamiltonians, from integrable Hamiltonians even in these
out of equilibrium situations.

It would be interesting to understand our results from a dual bulk
computation. We have already discussed this aspect in some detail in
the previous section. Here, we offer summarizing comments. An OTOC
computation in a thermal CFT translates to a geodesic computation in
a shockwave geometry created by a massive particle (dual to one of
the boundary operators), moving in a black hole
background\cite{Roberts:2014ifa}. Crucial to obtaining the
exponential time dependence is the existence of the black-hole
horizon, in the presence of which there is an exponentially large
blue shift in the energy of the massive particle.  The key
difference of the above-mentioned scenario with ours is that in our
case, OTOC is computed in the vacuum state. Under the $SL(2,R)$
drive, the vacuum state does not change. Thus the dual geometry,
would be AdS, instead of a black-hole. It would be interesting to
understand, how the exponential growth of the OTOC emerges from the
bulk in the absence of a black hole horizon. One possibility is that
the dual of the driven Hamiltonian generates the trajectory of an
accelerating particle and an effective horizon, similar to the
Rindler frame, emerges in this frame which is the source of the
exponential growth of the OTOC from the bulk. It would be nice if
this could be fleshed out in more detail.

%%%%%%%%%%%%%%%%%%%%%%%%%
\section*{~~~~~~~~~~~~~~~~~~~~~~~Acknowledgements}
%%%%%%%%%%%%%%%%%%%%%%%%%
We would like to thank Diptarka Das, Shouvik Datta for various conversations on related topics. SD would like to acknowledge the support provided by the Max Planck Partner Group grant MAXPLA/PHY/2018577.  BE is supported by CRG/2021/004539, AK acknowledges support from the Department of Atomic Energy, Govt.~of India, Board of Research in Nuclear Sciences (58/14/12/2021-BRNS)  and IFCPAR/CEFIPRA 6304-3. The work of SP and BR is supported by a Junior Research Fellowship(JRF) from UGC. KS thanks DST, India for support through project JCB/2021/000030.

\appendix
\numberwithin{equation}{section}
\section{OTOC in driven Ising CFT$_{2}$}
Here we briefly review the computation of OTOC in driven Ising CFTs.
We find that the large stroboscopic time behavior of OTOC in this
class of integrable systems remains similar to the large time
behaviour of the thermal OTOC \cite{Roberts:2014ifa}. Let us
consider 2d Ising CFT with Virasoro Identity operator $I$, spin
operator $\sigma$ and energy operator $\epsilon$. The four point
functions of these operators have explicit closed form expression as
follows:
\begin{align}
\frac{\langle \sigma\sigma\sigma\sigma \rangle}{\langle \sigma\sigma\rangle^{2}} &= \frac{1}{2}\left(\frac{1}{(1-\eta)(1-\bar{\eta})}\right)^{1/8} \left[\left(\sqrt{1+\sqrt{1-\eta}}\right)\left(\sqrt{1+\sqrt{1-\bar{\eta}}}\right) \right. \nonumber \\
&\left. +\left(\sqrt{1-\sqrt{1-\eta}}\right)\left(\sqrt{1-\sqrt{1-\bar{\eta}}}\right) \right] \ , \\
\frac{\langle \sigma\epsilon\sigma\epsilon \rangle}{\langle\sigma\sigma\rangle\langle\epsilon\epsilon\rangle} &= \frac{2-\eta}{2\sqrt{1-\eta}}\frac{2-\bar{\eta}}{2\sqrt{1-\bar{\eta}}} \ , \\
\frac{\langle \epsilon\epsilon\epsilon\epsilon \rangle}{\langle \epsilon\epsilon\rangle^{2}} &= \frac{1-\eta+\eta^{2}}{1-\eta}\frac{1-\bar{\eta}+\bar{\eta}^{2}}{1-\bar{\eta}} \ .
\end{align}
From the above expressions, we can see all these correlators have
branch cuts in the $\eta:[1,\infty)$ regime, except $\langle
\epsilon\epsilon\epsilon\epsilon \rangle$.  We have seen that for large values of the stroboscopic time:  $(\eta \sim
\epsilon_{12}\epsilon_{34}A(x)e^{-2n\theta})$\footnote{ We have a
similar form even for the general case when the operators are
located at distinct spatial points.}. Now, using
$(1-\eta)\rightarrow(1-\eta)e^{-2\pi i}$ and taking $\eta,\bar{\eta}
<< 1$ at large time, one gets:
\begin{align}
\frac{\langle \sigma\sigma\sigma\sigma \rangle}{\langle \sigma\sigma\rangle^{2}} \sim \frac{e^{i\pi/4}}{2}(\sqrt{\eta}+\sqrt{\bar{\eta}})+\mathcal{O}(\eta\bar{\eta}) \ .
\end{align}
Hence at large $n\theta$ it goes to zero. Similarly, we obtain for the other correlation functions:
\begin{align}
\frac{\langle \sigma\epsilon\sigma\epsilon \rangle}{\langle\sigma\sigma\rangle\langle\epsilon\epsilon\rangle}  &\sim -\left(1-\frac{\eta}{2}\right)\left(1-\frac{\bar{\eta}}{2}\right)\left(1+\frac{\eta}{2}\right)\left(1+\frac{\bar{\eta}}{2}\right) \nonumber \\
& \sim -1 + \mathcal{O}(\eta\bar{\eta},\dots) \ .
\end{align}
And for the  $\langle \epsilon\epsilon\epsilon\epsilon \rangle$ correlator we get:
\begin{equation}
\frac{\langle \epsilon\epsilon\epsilon\epsilon \rangle}{\langle \epsilon\epsilon\rangle^{2}} = 1 \ .
\end{equation}

Thus we get the usual behaviour of Ising OTOC where at late time these saturate to constant values ($0,-1$ and $1$ respectively). Thus even in driven CFT$_{2}$, OTOC distinguishes integrable systems from chaotic systems like large $c$ CFT$_{2}$.


\begin{thebibliography}{99}

\bibitem{rev1} J. Dziarmaga, Adv. Phys. {\bf 59}, 1063 (2010).

\bibitem{rev2} A. Polkovnikov, K. Sengupta, A. Silva, and M. Vengalattore,
Rev. Mod. Phys. {\bf 83}, 863 (2011).

\bibitem{rev3} A. Dutta, G. Aeppli, B. K. Chakrabarti, U. Divakaran,
T. F. Rosenbaum, and D. Sen, {\it Quantum phase transitions in
transverse field spin models: from statistical physics to quantum
information} (Cambridge University Press, Cambridge, 2015).

\bibitem{rev4} S. Mondal, D. Sen, and K. Sengupta, {\it Quantum Quenching,
Annealing and Computation}, edited by A. Das, A. Chandra, and B. K.
Chakrabarti, Lecture Notes in Physics, Vol. 802 (Springer, Berlin,
Heidelberg, 2010), Chap. 2, p. 21; C. De Grandi and A. Polkovnikov,
{\it ibid}, Chap 6, p. 75.

\bibitem{rev5}M. Bukov, L. D'Alessio and A. Polkovnikov, Advances
in Physics {\bf 64}, 139 (2015)

\bibitem{rev6} L. D'Alessio and A. Polkovnikov, Ann. Phys. {\bf 333}, 19 (2013).

\bibitem{rev7} L. D'Alessio, Y. Kafri, A. Polkovnikov, and M.
Rigol, Adv. Phys. {\bf 65}, 239 (2016).

\bibitem{rev8} A. Sen, D. Sen, and K. Sengupta, J. Phys. Condens. Matter {\bf 33},
443003 (2021).

\bibitem{rev9}I. Bloch, J. Dalibard, and W. Zwerger, Rev. Mod. Phys.
{\bf 80}, 885 (2008); L. Taurell and L. Sanchez-Palencia, C. R.
Physique {\bf 19}, 365 (2018).

\bibitem{exp1}M. Greiner, O. Mandel, T. Esslinger, T. W. Hansch, and I. Bloch,
Nature (London) {\bf 415}, 39 (2002); C. Orzel, A. K. Tuchman, M. L.
Fenselau, M. Yasuda, and M. A. Kasevich, Science {\bf 291}, 2386
(2001).

\bibitem{exp2}T. Kinoshita, T. Wenger, and D. S. Weiss, Nature
(London) {\bf 440}, 900 (2006); L. E. Sadler, J. M. Higbie, S. R.
Leslie, M. Vengalattore, and D. M. Stamper-Kurn, Nature (London)
{\bf 443}, 312 (2006).

\bibitem{exp3} W. Bakr, J. Gillen, A. Peng, S. Foelling and M. Greiner,
Nature {\bf 462}, 74 (2009); W. S. Bakr, A. Peng, M. E. Tai, R. Ma,
J. Simon, J. I. Gillen, S. Folling, L. Pollet, and M. Greiner,
Science {\bf 329}, 547 (2010).

\bibitem{exp4} H. Bernien, S. Schwartz, A. Keesling, H. Levine, A. Omran, H. Pichler,
S. Choi, A. S. Zibrov, M. Endres, M. Greiner, V. Vuletic, and M. D.
Lukin, Nature (London) {\bf 551}, 579 (2017); H. Levine, A.
Keesling, A. Omran, H. Bernien, S. Schwartz, A. S. Zibrov, M.
Endres, M. Greiner, V. Vuletic, and M. D. Lukin, Phys. Rev. Lett.
{\bf 121}, 123603 (2018).

\bibitem{magrev} S. Blanes, F. Casas, J.A. Oteo, and J. Ros,
Physics Reports {\bf 470}, 151 (2009).

\bibitem{fpt1} A. Soori and D. Sen, Phys. Rev. B {\bf 82}, 115432 (2010).

\bibitem{fpt2} T. Bilitewski and N. R. Cooper, Phys. Rev A {\bf 91}, 063611 (2015).

\bibitem{topo1}T. Kitagawa, E. Berg, M. Rudner, and E. Demler, Phys. Rev. B
{\bf 82}, 235114 (2010); N. H. Lindner, G. Refael, and V. Galitski,
Nat. Phys. {\bf 7}, 490 (2011); T. Kitagawa, T. Oka, A. Brataas, L.
Fu, and E. Demler, Phys. Rev. B {\bf 84}, 235108 (2011).

\bibitem{topo2} M. Thakurathi, A. A. Patel, D. Sen, and A. Dutta,
Phys. Rev. B {\bf 88}, 155133 (2013); A. Kundu, H. A. Fertig, and B.
Seradjeh, Phys. Rev. Lett. {\bf 113}, 236803 (2014).

\bibitem{topo3} F. Nathan and M. S. Rudner, New J. Phys. {\bf 17}, 125014 (2015); B.
Mukherjee, A. Sen, D. Sen, and K. Sengupta, Phys. Rev. B {\bf 94},
155122 (2016).

\bibitem{topo4}  B. Mukherjee, P. Mohan, D. Sen, and K. Sengupta, Phys. Rev. B
{\bf 97}, 205415 (2018).

\bibitem{tc1}V. Khemani, A. Lazarides, R.Moessner, and S. L. Sondhi, Phys.
Rev. Lett. {\bf 116}, 250401 (2016).

\bibitem{tc2}D. V. Else, B. Bauer, and C.
Nayak, Phys. Rev. Lett. {\bf 117}, 090402 (2016).

\bibitem{tc3} J. Zhang, P. W. Hess, A.
Kyprianidis, P. Becker, A. Lee, J. Smith, G. Pagano, I-D.
Potirniche, A. C. Potter, A. Vishwanath, N. Y. Yao, and C. Monroe,
Nature (London) {\bf 543}, 217 (2017).


\bibitem{dloc1} T. Nag, S. Roy, A. Dutta, and D. Sen, Phys. Rev. B {\bf 89}, 165425
(2014); T. Nag, D. Sen, and A. Dutta, Phys. Rev. A {\bf 91}, 063607
(2015).

\bibitem{dloc2} A. Agarwala, U. Bhattacharya, A. Dutta, and D. Sen, Phys. Rev. B
{\bf 93}, 174301 (2016); A. Agarwala and D. Sen, Phys. Rev. B {\bf
95}, 014305 (2017).

\bibitem{dloc3} D. J. Luitz, Y. Bar Lev, and A. Lazarides, SciPost Phys. {\bf 3}, 029
(2017); D. J. Luitz, A. Lazarides, and Y. Bar Lev, Phys. Rev. B {\bf
97}, 020303 (2018)

\bibitem{dloc4} R. Ghosh, B. Mukherjee, and K. Sengupta
Phys. Rev. B {\bf 102}, 235114(2020).

\bibitem{df1}A. Das, Phys.Rev. B {\bf 82}, 172402 (2010).

\bibitem{df2} S Bhattacharyya, A Das, and S Dasgupta, Phys. Rev. B {\bf 86} 054410 (2010).

\bibitem{df3} S. Hegde,H. Katiyar, T. S. Mahesh, and A. Das, Phys. Rev. B {\bf 90}, 174407
(2014)

\bibitem{df4}S. Mondal, D. Pekker, and K. Sengupta, Europhys. Lett.
{\bf 100}, 60007 (2012).

\bibitem{dtran1}M. Heyl, A. Polkovnikov, and S. Kehrein, Phys. Rev. Lett. {\bf 110},
135704 (2013); For a review, see M. Heyl, Rep. Prog. Phys {\bf 81},
054001 (2018).

\bibitem{dtran2}A. Sen, S. Nandy, and K. Sengupta, Phys. Rev. B {\bf 94}, 214301
(2016); S. Nandy, K. Sengupta, and A. Sen, J. Phys. A: Math. Theor.
{\bf 51}, 334002 (2018); M. Sarkar and K. Sengupta, Phys. Rev. B
{\bf 102}, 235154 (2020).

\bibitem{dtran3} S. Aditya, S. Samanta, A. Sen, K. Sengupta, and D. Sen,
arXiv:2112.02915 (unpublished).

\bibitem{dtran4} A. A. Makki, S. Bandyopadhyay, S. Maity, and A. Dutta,
arXiv:2112.02930 (unpublished).

\bibitem{erg1} B. Mukherjee, S. Nandy, A. Sen, D. Sen and K. Sengupta,
Phys. Rev B {\bf 101}, 245107 (2020).

\bibitem{erg2}B. Mukherjee, A. Sen, D. Sen
and K. Sengupta, Phys. Rev B {\bf 102}, 075123 (2020).

\bibitem{qd1}S. Nandy, A. Sen, and D. Sen, Phys. Rev. X {\bf 7}, 031034 (2017).

\bibitem{qd2}  A. Verdeny, J. Puig, and F. Mintert, Zeitschrift fur Naturforsch.
A {\bf 71}, 897 (2016); P. T. Dumitrescu, R. Vasseur, and A. C.
Potter, Phys. Rev. Lett. {\bf 120}, 070602 (2018).

\bibitem{qd3} S. Ray, S. Sinha, and D. Sen, Phys. Rev. E {\bf 100}, 052129
(2019); T. Mori, H. Zhao, F. Mintert, J. Knolle, and R. Moessner
Phys. Rev. Lett. {\bf 127}, 050602 (2021).

\bibitem{nd1} H. Zhao, F. Mintert, R. Moessner, and J. Knolle, Phys. Rev. Lett.
{\bf 126}, 040601 (2021); B. Mukherjee, A. Sen, D. Sen, and K.
Sengupta, Phys. Rev. B {\bf 102}, 014301 (2020).

%\cite{Nizami:2020agu}
\bibitem{Nizami:2020agu}
A.~A.~Nizami,
%``Quantum chaos measures for Floquet dynamics,''
[arXiv:2007.07283 [quant-ph]].
%1 citations counted in INSPIRE as of 07 Apr 2022

\bibitem{sdas1} S. R. Das, D. A. Galante, and R. C. Myers, Phys. Rev. Lett.
{\bf 112}, 171601 (2014).

%\cite{Wen:2018agb}
\bibitem{Wen:2018agb}
X.~Wen and J.~Q.~Wu,
%``Floquet conformal field theory,''
[arXiv:1805.00031 [cond-mat.str-el]].
%28 citations counted in INSPIRE as of 19 Nov 2021


%\cite{Wen:2020wee}
\bibitem{Wen:2020wee}
X.~Wen, R.~Fan, A.~Vishwanath and Y.~Gu,
%``Periodically, quasiperiodically, and randomly driven conformal field theories,''
Phys. Rev. Res. \textbf{3}, no.2, 023044 (2021)
doi:10.1103/PhysRevResearch.3.023044
[arXiv:2006.10072 [cond-mat.stat-mech]].
%13 citations counted in INSPIRE as of 19 Nov 2021


%\cite{Han:2020kwp}
\bibitem{Han:2020kwp}
B.~Han and X.~Wen,
%``Classification of $SL_2$ deformed Floquet conformal field theories,''
Phys. Rev. B \textbf{102}, no.20, 205125 (2020)
doi:10.1103/PhysRevB.102.205125
[arXiv:2008.01123 [cond-mat.stat-mech]].
%7 citations counted in INSPIRE as of 19 Nov 2021


%\cite{Fan:2020orx}
\bibitem{Fan:2020orx}
R.~Fan, Y.~Gu, A.~Vishwanath and X.~Wen,
%``Floquet conformal field theories with generally deformed Hamiltonians,''
SciPost Phys. \textbf{10}, no.2, 049 (2021)
doi:10.21468/SciPostPhys.10.2.049 [arXiv:2011.09491 [hep-th]].
%6 citations counted in INSPIRE as of 20 Nov 2021
%\cite{Andersen:2020xvu}

\bibitem{Andersen:2020xvu}
M.~Andersen, F.~N\o{}rfjand and N.~T.~Zinner,
%``Real-time correlation function of Floquet conformal fields,''
Phys. Rev. D \textbf{103}, no.5, 056005 (2021)
doi:10.1103/PhysRevD.103.056005
[arXiv:2011.08494 [cond-mat.str-el]].
%4 citations counted in INSPIRE as of 19 Nov 2021


%\cite{Das:2021gts}
\bibitem{Das:2021gts}
D.~Das, R.~Ghosh and K.~Sengupta,
%``Conformal Floquet dynamics with a continuous drive protocol,''
JHEP \textbf{05} (2021), 172
doi:10.1007/JHEP05(2021)172
[arXiv:2101.04140 [hep-th]].
%2 citations counted in INSPIRE as of 18 Nov 2021


%\cite{Lapierre:2019rwj}
\bibitem{Lapierre:2019rwj}
B.~Lapierre, K.~Choo, C.~Tauber, A.~Tiwari, T.~Neupert and R.~Chitra,
%``Emergent black hole dynamics in critical Floquet systems,''
Phys. Rev. Res. \textbf{2}, no.2, 023085 (2020)
doi:10.1103/PhysRevResearch.2.023085
[arXiv:1909.08618 [cond-mat.str-el]].
%14 citations counted in INSPIRE as of 07 Apr 2022


%\cite{Lapierre:2020roc}
\bibitem{Lapierre:2020roc}
B.~Lapierre, K.~Choo, A.~Tiwari, C.~Tauber, T.~Neupert and R.~Chitra,
%``Fine structure of heating in a quasiperiodically driven critical quantum system,''
Phys. Rev. Res. \textbf{2}, no.3, 033461 (2020)
doi:10.1103/PhysRevResearch.2.033461
%10 citations counted in INSPIRE as of 07 Apr 2022


%\cite{Lapierre:2020ftq}
\bibitem{Lapierre:2020ftq}
B.~Lapierre and P.~Moosavi,
%``Geometric approach to inhomogeneous Floquet systems,''
Phys. Rev. B \textbf{103}, 224303 (2021)
doi:10.1103/PhysRevB.103.224303
[arXiv:2010.11268 [cond-mat.stat-mech]].
%8 citations counted in INSPIRE as of 07 Apr 2022


%\cite{Okunishi:2016zat}
\bibitem{Okunishi:2016zat}
K.~Okunishi,
%``Sine-square deformation and M\"obius quantization of 2D conformal field theory,''
PTEP \textbf{2016}, no.6, 063A02 (2016) doi:10.1093/ptep/ptw060
[arXiv:1603.09543 [hep-th]].
%20 citations counted in INSPIRE as of 19 Nov 2021


%\cite{Milsted:2017csn}
\bibitem{Milsted:2017csn}
A.~Milsted and G.~Vidal,
%``Extraction of conformal data in critical quantum spin chains using the Koo-Saleur formula,''
Phys. Rev. B \textbf{96}, no.24, 245105 (2017)
doi:10.1103/PhysRevB.96.245105 [arXiv:1706.01436 [cond-mat.str-el]].
%34 citations counted in INSPIRE as of 19 Nov 2021


\bibitem{Das:2019tga}
S.~Das, B.~Ezhuthachan and A.~Kundu,
%``Real time dynamics from low point correlators in 2d BCFT,''
JHEP \textbf{12}, 141 (2019)

%\cite{Dolan:2000ut}
\bibitem{Dolan:2000ut}
F.~A.~Dolan and H.~Osborn,
%``Conformal four point functions and the operator product expansion,''
Nucl. Phys. B \textbf{599}, 459-496 (2001)
doi:10.1016/S0550-3213(01)00013-X
[arXiv:hep-th/0011040 [hep-th]].
%508 citations counted in INSPIRE as of 08 Nov 2021

%\cite{Fitzpatrick:2014vua}
\bibitem{Fitzpatrick:2014vua}
A.~L.~Fitzpatrick, J.~Kaplan and M.~T.~Walters,
%``Universality of Long-Distance AdS Physics from the CFT Bootstrap,''
JHEP \textbf{08}, 145 (2014)
doi:10.1007/JHEP08(2014)145
[arXiv:1403.6829 [hep-th]].

%\cite{Roberts:2014ifa}
\bibitem{Roberts:2014ifa}
D.~A.~Roberts and D.~Stanford,
%``Two-dimensional conformal field theory and the butterfly effect,''
Phys. Rev. Lett. \textbf{115}, no.13, 131603 (2015)
doi:10.1103/PhysRevLett.115.131603
[arXiv:1412.5123 [hep-th]].
%297 citations counted in INSPIRE as of 08 Nov 2021


%\cite{Das:2021qsd}
\bibitem{Das:2021qsd}
S.~Das, B.~Ezhuthachan, A.~Kundu, S.~Porey and B.~Roy,
%``Critical Quenches, OTOCs and Early-Time Chaos,''
[arXiv:2108.12884 [hep-th]].

\bibitem{Aramthottil:2021cov}
A.~S.~Aramthottil, D.~Das, S.~Das and B.~Dey,
%``Scrambling under quench,''
[arXiv:2109.02132 [hep-th]].

\bibitem{otoc1} For a review see, S. Xu and B. Swingle,
arXiv:2202.07060.

\bibitem{otoc2a}J. Mumford, W. Kirkby, and D. H. J. O'Dell,
J. Phys. B: At. Mol. Opt. Phys. {\bf 53}, 145301 (2020)

\bibitem{otoc2b} S. Ray, S. Sinha, and K. Sengupta,
Phys. Rev. A {\bf 98}, 053631 (2018).

%\cite{Maldacena:2015waa}
\bibitem{Maldacena:2015waa}
J.~Maldacena, S.~H.~Shenker and D.~Stanford,
%``A bound on chaos,''
JHEP \textbf{08}, 106 (2016)
doi:10.1007/JHEP08(2016)106
[arXiv:1503.01409 [hep-th]].
%1219 citations counted in INSPIRE as of 12 Feb 2022


%\cite{Lieb:1972wy}
\bibitem{Lieb:1972wy}
E.~H.~Lieb and D.~W.~Robinson,
%``The finite group velocity of quantum spin systems,''
Commun. Math. Phys. \textbf{28}, 251-257 (1972)
doi:10.1007/BF01645779
%411 citations counted in INSPIRE as of 12 Feb 2022



%\cite{Xu:2018dfp}
\bibitem{Xu:2018dfp}
S.~Xu and B.~Swingle,
%``Locality, Quantum Fluctuations, and Scrambling,''
Phys. Rev. X \textbf{9}, no.3, 031048 (2019)
doi:10.1103/PhysRevX.9.031048
[arXiv:1805.05376 [cond-mat.str-el]].
%66 citations counted in INSPIRE as of 12 Feb 2022


%\cite{Gharibyan:2018fax}
\bibitem{Gharibyan:2018fax}
H.~Gharibyan, M.~Hanada, B.~Swingle and M.~Tezuka,
%``Quantum Lyapunov Spectrum,''
JHEP \textbf{04}, 082 (2019)
doi:10.1007/JHEP04(2019)082
[arXiv:1809.01671 [quant-ph]].
%20 citations counted in INSPIRE as of 12 Feb 2022


%\cite{Steinberg:2019uqb}
\bibitem{Steinberg:2019uqb}
J.~Steinberg and B.~Swingle,
%``Thermalization and chaos in QED$_{3}$,''
Phys. Rev. D \textbf{99}, no.7, 076007 (2019)
doi:10.1103/PhysRevD.99.076007
[arXiv:1901.04984 [cond-mat.str-el]].
%12 citations counted in INSPIRE as of 12 Feb 2022

%\cite{Gu:2021xaj}
\bibitem{Gu:2021xaj}
Y.~Gu, A.~Kitaev and P.~Zhang,
%``A two-way approach to out-of-time-order correlators,''
[arXiv:2111.12007 [hep-th]].
%4 citations counted in INSPIRE as of 12 Feb 2022


%\cite{Poojary:2018esz}
\bibitem{Poojary:2018esz}
R.~R.~Poojary,
%``BTZ dynamics and chaos,''
JHEP \textbf{03}, 048 (2020)
doi:10.1007/JHEP03(2020)048
[arXiv:1812.10073 [hep-th]].
%28 citations counted in INSPIRE as of 12 Feb 2022


%\cite{Banerjee:2018twd}
\bibitem{Banerjee:2018twd}
A.~Banerjee, A.~Kundu and R.~R.~Poojary,
%``Strings, Branes, Schwarzian Action and Maximal Chaos,''
[arXiv:1809.02090 [hep-th]].
%20 citations counted in INSPIRE as of 12 Feb 2022


%\cite{Malvimat:2021itk}
\bibitem{Malvimat:2021itk}
V.~Malvimat and R.~R.~Poojary,
%``Fast Scrambling due to Rotating Shockwaves in BTZ,''
[arXiv:2112.14089 [hep-th]].
%0 citations counted in INSPIRE as of 12 Feb 2022


%\cite{Rangamani:2015sha}
\bibitem{Rangamani:2015sha}
M.~Rangamani, M.~Rozali and A.~Wong,
%``Driven Holographic CFTs,''
JHEP \textbf{04}, 093 (2015)
doi:10.1007/JHEP04(2015)093
[arXiv:1502.05726 [hep-th]].
%24 citations counted in INSPIRE as of 11 Feb 2022


%\cite{Biasi:2017kkn}
\bibitem{Biasi:2017kkn}
A.~Biasi, P.~Carracedo, J.~Mas, D.~Musso and A.~Serantes,
%``Floquet Scalar Dynamics in Global AdS,''
JHEP \textbf{04}, 137 (2018)
doi:10.1007/JHEP04(2018)137
[arXiv:1712.07637 [hep-th]].
%18 citations counted in INSPIRE as of 07 Apr 2022


%\cite{Biasi:2019eap}
\bibitem{Biasi:2019eap}
A.~Biasi, J.~Mas and A.~Serantes,
%``Gravitational wave driving of a gapped holographic system,''
JHEP \textbf{05}, 161 (2019)
doi:10.1007/JHEP05(2019)161
[arXiv:1903.05618 [hep-th]].
%12 citations counted in INSPIRE as of 07 Apr 2022


%\cite{Banerjee:2017lhg}
\bibitem{Banerjee:2017lhg}
A.~Banerjee, A.~Kundu, P.~Roy and A.~Virmani,
%``Oscillating Shells and Oscillating Balls in AdS,''
JHEP \textbf{07}, 026 (2017)
doi:10.1007/JHEP07(2017)026
[arXiv:1704.07570 [hep-th]].
%1 citations counted in INSPIRE as of 12 Feb 2022


%\cite{Berenstein:2019tcs}
\bibitem{Berenstein:2019tcs}
D.~Berenstein and J.~Sim\'on,
%``Localized states in global AdS space,''
Phys. Rev. D \textbf{101}, no.4, 046026 (2020)
doi:10.1103/PhysRevD.101.046026
[arXiv:1910.10227 [hep-th]].
%4 citations counted in INSPIRE as of 12 Feb 2022


%\cite{Banerjee:2019vff}
\bibitem{Banerjee:2019vff}
A.~Banerjee, A.~Kundu and R.~R.~Poojary,
%``Rotating black holes in AdS spacetime, extremality, and chaos,''
Phys. Rev. D \textbf{102}, no.10, 106013 (2020)
doi:10.1103/PhysRevD.102.106013
[arXiv:1912.12996 [hep-th]].
%10 citations counted in INSPIRE as of 12 Feb 2022


%\cite{Craps:2020ahu}
\bibitem{Craps:2020ahu}
B.~Craps, M.~De Clerck, P.~Hacker, K.~Nguyen and C.~Rabideau,
%``Slow scrambling in extremal BTZ and microstate geometries,''
JHEP \textbf{03}, 020 (2021)
doi:10.1007/JHEP03(2021)020
[arXiv:2009.08518 [hep-th]].
%15 citations counted in INSPIRE as of 12 Feb 2022


\bibitem{Gu:2016hoy}
Y.~Gu and X.~L.~Qi,
%``Fractional Statistics and the Butterfly Effect,''
JHEP \textbf{08}, 129 (2016)

\bibitem{Caputa:2016tgt}
P.~Caputa, T.~Numasawa and A.~Veliz-Osorio,
%``Out-of-time-ordered correlators and purity in rational conformal field theories,''
PTEP \textbf{2016}, no.11, 113B06 (2016)
%\cite{Fan:2018ddo}
\bibitem{Fan:2018ddo}
R.~Fan,
%``Out-of-Time-Order Correlation Functions for Unitary Minimal Models,''
[arXiv:1809.07228 [hep-th]].























%\bibitem{Sotiriadis:2008ila}
%S.~Sotiriadis and J.~Cardy,
%``Inhomogeneous Quantum Quenches,''
%J. Stat. Mech. \textbf{0811}, P11003 (2008)


%\bibitem{Calabrese:2006rx}
%P.~Calabrese and J.~L.~Cardy,
%``Time-dependence of correlation functions following a quantum quench,''
%Phys. Rev. Lett. \textbf{96}, 136801 (2006)


%\bibitem{Calabrese:2007rg}
%P.~Calabrese and J.~Cardy,
%``Quantum Quenches in Extended Systems,''
%J. Stat. Mech. \textbf{0706}, P06008 (2007)


%\bibitem{Cardy:2015xaa}
%J.~Cardy,
%``Quantum Quenches to a Critical Point in One Dimension: some further results,''
%J. Stat. Mech. \textbf{1602}, no.2, 023103 (2016)


%\bibitem{Calabrese:2016xau}
%P.~Calabrese and J.~Cardy,
%``Quantum quenches in 1  +  1 dimensional conformal field theories,''
%J. Stat. Mech. \textbf{1606}, no.6, 064003 (2016)


%\bibitem{Das:2019tga}
%S.~Das, B.~Ezhuthachan and A.~Kundu,
%``Real time dynamics from low point correlators in 2d BCFT,''
%JHEP \textbf{12}, 141 (2019)


%\bibitem{Roberts:2014ifa}
 % D.~A.~Roberts and D.~Stanford,
 % ``Two-dimensional conformal field theory and the butterfly effect,''
 % Phys.\ Rev.\ Lett.\  {\bf 115}, no. 13, 131603 (2015)
 % doi:10.1103/PhysRevLett.115.131603
 % [arXiv:1412.5123 [hep-th]].


\end{thebibliography}
\end{document}